\documentclass[pre,pr1,reprint,twocolumn,amsmath,10pt,amssymb,aps,superscriptaddress,showkeys,showpacs]{revtex4-1}
\usepackage[utf8]{inputenc}
\usepackage{lipsum}
\usepackage{enumitem,mathtools}
\usepackage{graphicx}
\usepackage{subcaption}
\usepackage{dcolumn}
\usepackage{array}
\usepackage{bm}
\usepackage[colorlinks=true,
            linkcolor=blue,
            urlcolor  = blue,
            citecolor = blue,
            anchorcolor = blue]{hyperref}
\usepackage[
scale=0.7, marginratio={1:1, 2:3}, ignoreall,
text={7in,10in},centering,
margin=.6in,
total={6.5in,8.75in}, top=1.2in, left=0.5in, includefoot,
height=10in,a4paper,hmargin={1.5cm,0.8in},
]{geometry}
\usepackage{fancyhdr}
\usepackage{mwe}
\usepackage{relsize}
\usepackage{scalerel}
\usepackage{xparse}
\usepackage{array} 
\usepackage[utf8]{inputenc}
\newcommand{\be}{\begin{equation}}
\newcommand{\bea}{\begin{align}}
\newcommand{\ee}{\end{equation}}
\newcommand{\eea}{\end{align}}
\usepackage{etoolbox} 
\usepackage{xcolor}
\usepackage[capitalize]{cleveref}

\makeatletter
\appto{\appendix}{%
  \@ifstar{\def\thesection{\unskip}\def\theequation@prefix{A.}}%
          {\def\thesection{\Alph {section}}}%
}
\makeatother
\begin{document}
\title{First passage in the presence of stochastic resetting and a potential  barrier}
\author{Saeed Ahmad}
\email{saeedmalik@iitb.ac.in}%
\author{Krishna Rijal}%
\author{Dibyendu Das}
\email{dibyendu@phy.iitb.ac.in}%
\affiliation{Physics Department, Indian Institute of Technology Bombay, Mumbai 400076, India}%
\date{\today}
\begin{abstract}
    Diffusion and first passage in the presence of stochastic resetting and potential bias have been of recent interest. We study a few models, systematically progressing in their complexity, to understand the usefulness of resetting. In the parameter space of the models, there are multiple continuous and discontinuous transitions where the advantage of resetting vanishes. We show these results analytically exactly for a tent-potential, and numerically accurately for a quartic-potential relevant to a magnetic system at low temperatures. We find that the spatial asymmetry of the potential across the barrier, and the number of absorbing boundaries, play a crucial role in determining the type of transition.

\begin{description}
\item[PACS number(s)]05.40.-a,05.70.Fh,02.50.-r,02.50.Ey
\end{description}
\end{abstract}
\maketitle

\section{\label{sec1}Introduction} 
Problems of diffusion with stochastic resetting has become an important field of study in recent years~\cite{PRL_Evan_Staya_2011,Review_Evans_Satya_Reset_application_2019}. In any stochastic process, the event of resetting instantly sets the system back to its initial state after random times. These time intervals are drawn from distributions which may have various forms~\cite{PRL_Evan_Staya_2011,Evan_Staya_2011Position,J_Stat_Mech_determinitic_Bhat_2016,PRL_Reuveni_2016First,PRL_APAl_Reuveni_2017FirstR, PRE_pwer_law_rate_Apoorva_ShamikG_2016,J_Phys_A_ApalKundu_Evan2016reset_t,PRE_Shilkev_2017_ContiRand_R-t}. Non-instantaneous resetting has also been studied~\cite{PRE_Non_instantaneous_reset_diffudion_Sokolov_2020}. The idea of resetting has found applications in the problems of chemical reactions~\cite{PNAS_MicMenten_Reuveni_Shlomi_2014,PRE_MM_unified_Rotbart_Reuveni_2015,Nat_comm_Single_molecule_Robin_Reuveni_Urbakh_2018}, biological processes~\cite{PRE_backtrac_RNA_polymer_Roldan_2016,PRE_mutiple_target_Bressloff_2020,J_Phys_A_partial_ab_Bresloff_Schumm_2021}, and magnetic phase transitions~\cite{PRR_Ising_model_w_reset_staya_m_2020}. Resetting in the presence of space dependent potentials introduce interesting new features~\cite{PRE_Potential_APal_2015,J_Phys_A_Reset_Boun_Christo_2015,PRE_rx_harmonic_pot_Roldan_Shamik_2017,PRE_Saeed_2019,J_Phys_A_Peclet_n_Ray_Debasish_Reuveni_2019,PRR_Pal_Parsad_Landau_2019,J_Chem_Phys_log_Pot_Reuveni_2020,PRE_Saeed_Das_2020,J_Phys_A_reset_confining_pot_Metzler_Singh_2020,J_Chem_Phys_dble_w_pot_temp_Reuveni_Ray_2021}. Our work below is related to the latter theme.

One of the motivations to study resetting comes from the fact that it may be used as a strategy to expedite the first passage to a target~\cite{PRL_Evan_Staya_2011}. It is now known that for resetting at a constant rate, the optimal resetting rate (ORR) coincides with the condition $CV=1$, where the noise $CV$ is the ratio of standard deviation to mean of first passage time~\cite{PRL_Reuveni_2016First}. The optimality for other resetting time distributions were also studied~\cite{PRL_APAl_Reuveni_2017FirstR}. First passage to a target on the other hand can also be regulated by tuning the spatial potential under which  diffusion happens. When both potential bias and resetting work together, the former may render the latter strategy irrelevant at a point --- this leads to transitions between phases with or without benefit of resetting~\cite{PRE_Saeed_2019,J_Phys_A_Reset_Boun_Christo_2015,J_Phys_A_Peclet_n_Ray_Debasish_Reuveni_2019,PRR_Pal_Parsad_Landau_2019,J_Chem_Phys_log_Pot_Reuveni_2020,PRE_Saeed_Das_2020,J_Phys_A_reset_confining_pot_Metzler_Singh_2020,J_Chem_Phys_dble_w_pot_temp_Reuveni_Ray_2021}.

For generic spherically symmetric potentials $V(R)=kR^{n}$ and $k\ln(R/a)$ in any dimension $d$, the continuous transition point $k_{c}$ where ORR vanishes  was exactly solved for various cases~\cite{PRE_Saeed_Das_2020} --- it was shown that $k_c \sim dR^{-n}_{0}$ for the power-law potential (where $R_0$ is the resetting radius) and $k_c \sim d$ for the logarithimic potential, at large $d$. Various studies showed that the optimal resetting rate vanishing transition (ORRVT) has similarities to usual phase transitions --- the mean first passage time (MFPT) is analogous to free energy and ORR is analogous to an order parameter~\cite{PRE_Saeed_2019,PRR_Pal_Parsad_Landau_2019, PRE_Saeed_Das_2020}. A Landau like theory for MFPT as a function of ORR was developed~\cite{PRE_Saeed_2019,PRR_Pal_Parsad_Landau_2019}. In particular, the possibility of discontinuous transitions of ORR, and continuous tri-critical point (TCP) was demonstrated in~\cite{PRR_Pal_Parsad_Landau_2019}. A natural question is which factors influence the nature of the transition, i.e continuous or discontinuous. The models that we study in this paper show that increase in the number of absorbing boundaries play a crucial role in giving rise to discontinuous transitions. This is reminiscent of switch from continuous to discontinuous transition in equilibrium $q$-state Potts model with increasing $q$~\cite{Rev_Mod_Phys_Potts_model_1982}. 

In the context of a barrier crossing problem~\cite{Gardiner_book_2004, PRL_Freezing_transition_barrier_c_sajib_staya_2020} relevant to chemical reactions, the usefulness of resetting and associated ORRVT was studied in~\cite{PRE_Saeed_2019,J_Phys_A_reset_confining_pot_Metzler_Singh_2020,J_Chem_Phys_dble_w_pot_temp_Reuveni_Ray_2021}. The latter studies focused on the question of the first passage to a hill top, starting from an adjacent valley. Yet if one considers a magnetic system in which the magnetization $+m_0$ has to evolve to $-m_0$, the initial state and the final target are both at valleys seperated by a potential barrier which is to be crossed. Such a problem with resetting has not been studied earlier. We study this problem analytically in this paper using various models. One model deals with a Brownian particle in a piecewise linear potential which is tent-like, while another model deals with a magnetic system evolving at a fixed temperature and magnetic field under a Landau-like quartic potential. We show that the left-right asymmetry of the potential wells in these models, play a crucial role in determining the number of transitions in the parameter space and the type of transition.

This article is organized as follows. In section 2, we define four models based on the shapes of the potentials and the type of boundary conditions. In section 3, we discuss the methods used to study the transitions of ORR. In section 4, we present our results on ORRVT in two models in which only continuous transition (CT) arise. In section 5, we study the other two models where discontinuous transition (DT) and TCPs arise.  Finally, we conclude in section 6.

 \section{The Models}
\begin{figure}[ht!]
  \begin{minipage}{.5\textwidth}
    \begin{subfigure}[b]{0.45\textwidth}
      \includegraphics[width = 1.0\textwidth,height=0.1\textheight]{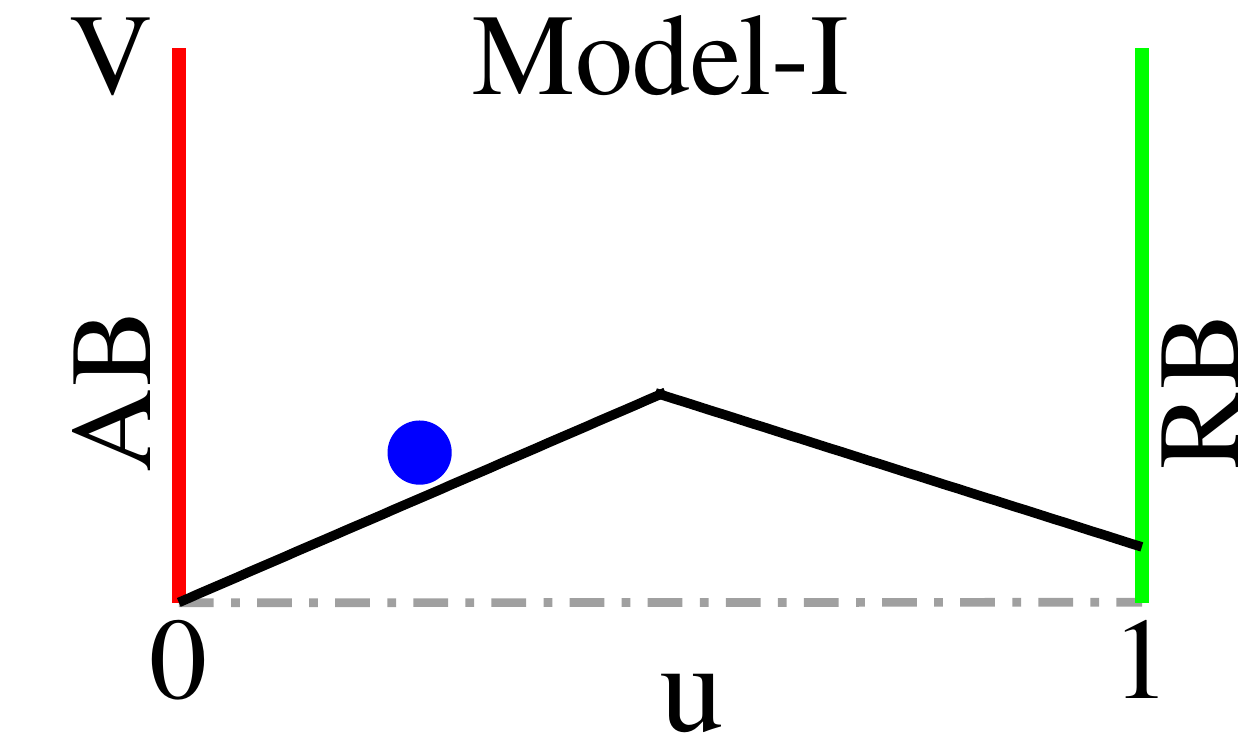}
      \caption{}\label{fig:AB_RB_tent_pot}
    \end{subfigure}
    \begin{subfigure}[b]{0.45\textwidth}
      \includegraphics[width = 1.0\textwidth,height=0.1\textheight]{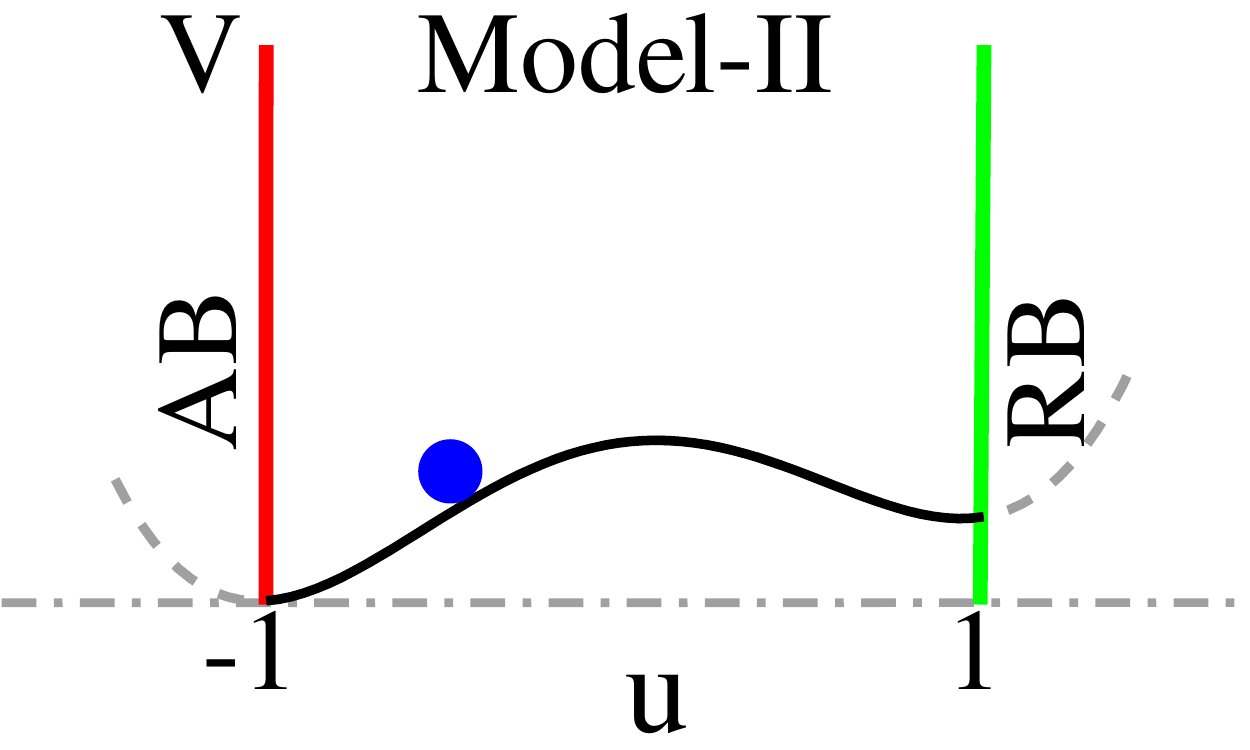}
      \caption{}\label{fig:AB_RB_Ising_pot}
    \end{subfigure}
    \begin{subfigure}[b]{0.45\textwidth}
        \includegraphics[width = 1.0\textwidth,height=0.1\textheight]{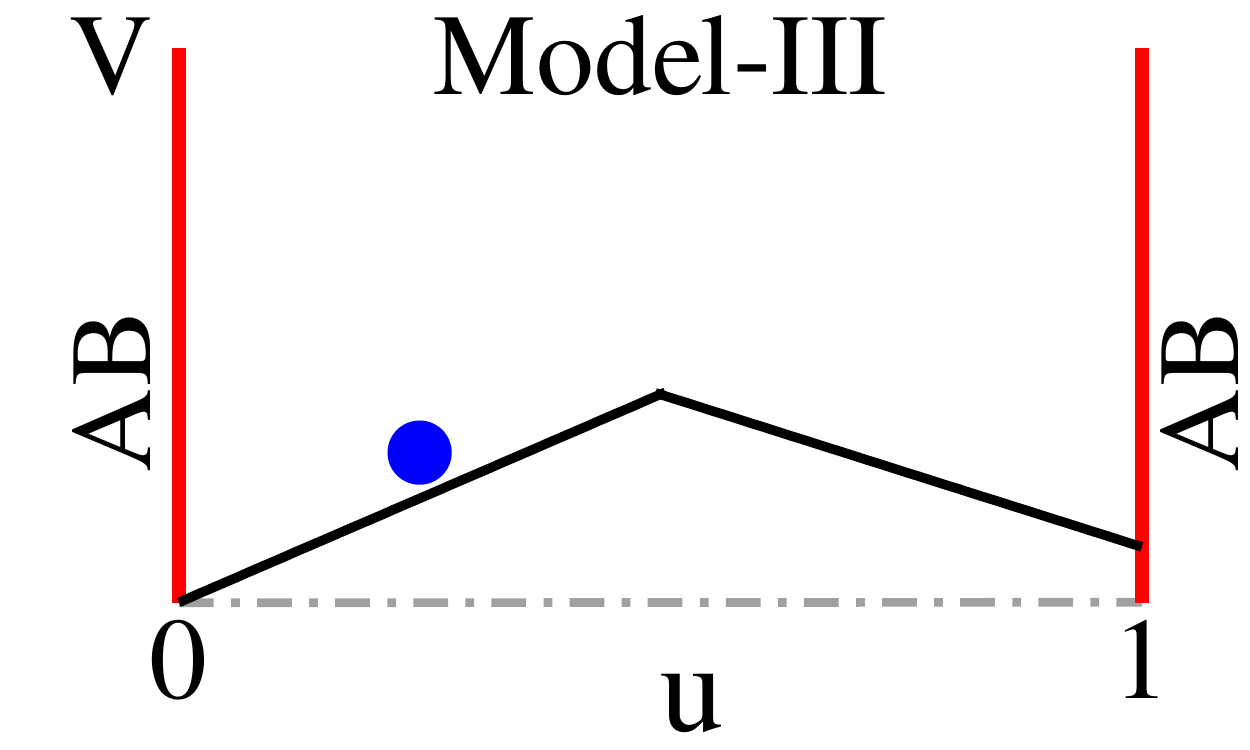}
        \caption{}
        \label{fig:AB_AB_tent_pot}
    \end{subfigure}
    \begin{subfigure}[b]{0.45\textwidth}
      \includegraphics[width = 1.0\textwidth,height=0.1\textheight]{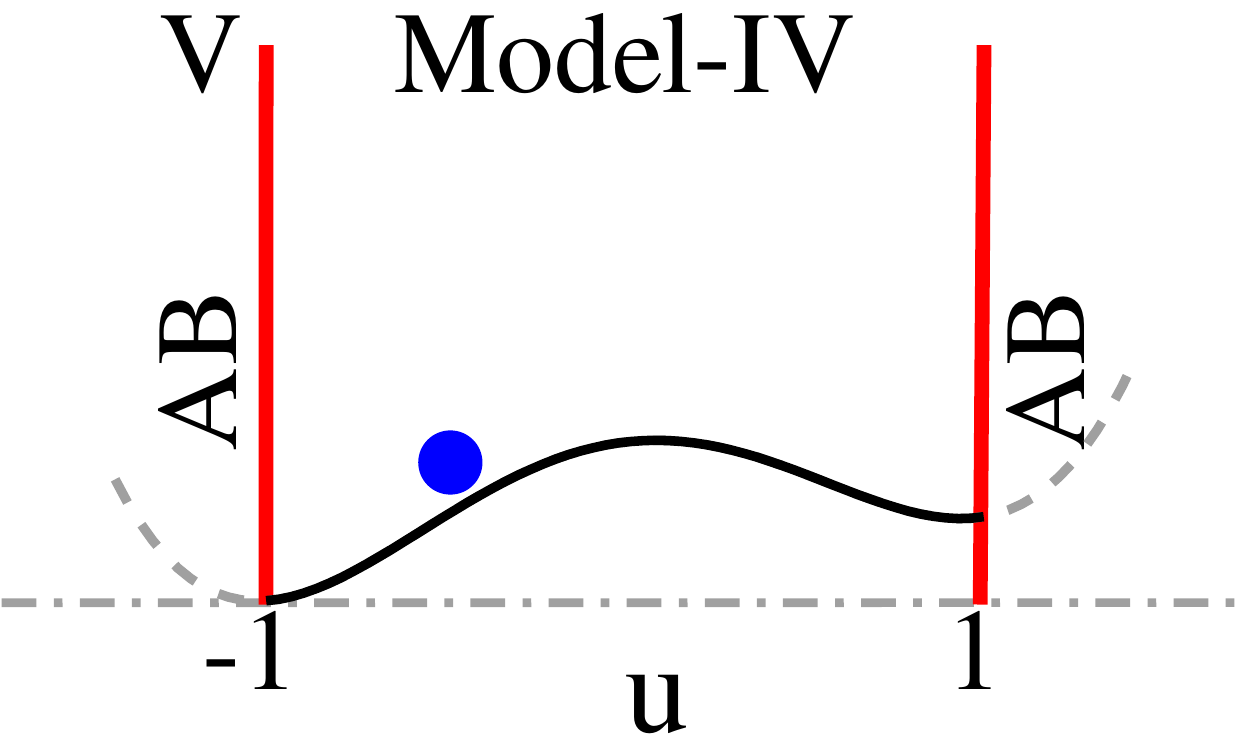}
      \caption{}\label{fig:AB_AB_Ising_pot}
    \end{subfigure}
\caption{The four models discussed in the text are schematically shown here with the potential $V$ plotted as a function of scaled distance (or magnetization) $u$ --- for (a) Model-I and (c) Model-III, $u=x/L$ while for (b) Model-II and (d) Model-IV,  $u=\frac{x\sqrt{2}}{\sqrt{b}}$. In all the models the left boundary is an AB. In (a) and (b) the right boundary is a RB, while in (c) and (d) right boundary is an absorbing one.}
    \label{fig:All_model}
  \end{minipage}
  \hfill
  \begin{minipage}{0.4\textwidth}
  \end{minipage}
\end{figure}

In this article, we explore the role of the shape of the potential and the boundary conditions which give rise to rich diversity of transitions demarcating the phases where resetting is beneficial or a hindrance. We study four different models in this paper, which are shown in Fig.(\ref{fig:All_model}). In all these models the potential has a non-monotonic shape with a hill in between two valleys --- a new feature in comparison to the earlier studies.

In Model-I and Model-III (Fig.(\ref{fig:AB_RB_tent_pot}, \ref{fig:AB_AB_tent_pot})) we have a diffusing particle subjected to a tent potential $V(x)$ over $x\in [0,L]$ given by the Eq.(\ref{Tent_pot}):
\be
\mathrm{V}(x) =
\begin{cases}
 \hfill  v_1(x-x_m) ,&  \text{$x\leq x_m$},\\
 \hfill -v_2(x-x_m) & \text{$x\geq x_m$}.
\end{cases}
\label{Tent_pot}
\ee
where the linear segments of slopes $v_1$ and $-v_2$ are separated by a peak at $x_m$. The values of the slopes determine the depth of the valleys formed at the boundaries $u=0$ and at $u=1$ (where $u$ is the scaled length $x/L$). The particle position is stochastically reset to its initial position $x_0\in (0,L)$ at a constant rate $r$. Since our main attention is to study the first passage problems we have to suitably define the target as an absorbing boundary (AB). In Model-I the left boundary at $u=0$ is an AB while the right boundary at $u=1$ is a reflecting boundary (RB) (i.e $V(x)=\infty$ for $x\geq L$). In contrast in Model-III we have two targets at $u=0$ and $u=1$ both of which serve as AB --- the first passage is achieved when the particle arrives at any one of them for the first time. We would see below that the seemingly innocuous change of the nature of the right boundary at $u=1$ makes the Model-III have a completely different phase diagram compared to Model-I. We note that for $v_1<0$ and $v_2<0$ (i.e. a valley at $x_m$) with no RB but only an AB, our Model-I becomes identical to the one studied in~\cite{ J_Phys_A_reset_confining_pot_Metzler_Singh_2020}.

In Model-II and Model-IV the position $x$ of the diffusing particle represents the magnetization state of a magnetic system. The potential $V(x)$ has a mean field Landau form given by
\be
V(x)=x^4-bx^2-cx
\label{Landau_phi_pot}
\ee
where $b=\frac{6(Tc-T)}{T_c}$ represents the reduced temperature (deviation from the critical temperature $T_c$) and $c=\frac{12h}{T_c}$ is the reduced magnetic field. In the symmetric case of $c=0$ (zero external field) the minima of the potential are at $x=\pm\sqrt{\frac{b}{2}}$ and the maximum is at $x=0$. We define a scaled magnetization $u=\frac{x\sqrt{2}}{\sqrt{b}}$. In all our studies the system is confined to the domain $u\in [-1,1]$, even when the potential is asymmetric ($|c|>0$). Note that  in the asymmetric case ($|c|>0$) the maximum is shifted from the location $u=0$ and the minima are not at $u=\pm 1$ --- instead their locations may be solved from the cubic algebraic equation: $V^{'}(x)=4x^3-2bx-c=0$. The system is stochastically reset to its initial magnetization $u_0\in(-1,1)$ at a constant rate $r$. In Model-II the target for the first passage is the negative magnetization state $u_0=-1$ achieved by an AB condition while $u_0=1$ serves as RB. In Model-IV both positive and negative magnetization states $u_0=\pm 1$ are ABs and serve as targets for the first passage. In both the models it would be interesting to see whether resetting strategy helps to reach the target magnetization state(s) at various values of the reduced temperature $b$ and magnetic field $c$. Note that with increasing $b$ (lower temperatures) the barrier height increases making the first passage more challenging. In such situations the asymmetry introduced by the magnetic field as well as stochastic resetting may help reach the target(s) --- explicit study below will show where the field renders the resetting strategy ineffective.

For all the four models the stochastic process is described by the forward Chapman-Kolmogorov equation
\be
\frac{\partial P}{\partial t}=
D\frac{\partial^{2} P}{\partial x^{2}}+\frac{\partial [V^{\prime}(x) P]}{\partial x}-rP+r\delta(x-x_0),  
\label{eq:Farward_FP}
\ee 
where $P\equiv P(x,t|x_0)$ is the probability of finding the particle at position (or magnetization) $x$ at time $t$ starting from $x_0$ at time $t=0$ i.e. with initial condition $P(x,0|x_0)=\delta(x-x_0)$. Here $r$ is constant rate of resetting to the initial position $x_0$ and $D$ is the diffusion constant. In addition, there are specific boundary conditions relevant to the different models. In this paper, we would not be using the forward formalism, but the backward stochastic  formalism which is more suitable for obtaining the properties related to the first passage. In the following, we would discuss the backward master equation for survival probability and the general method of finding ORRVT for the models.


\section{\label{sec2}Method to obtain ORRVT for the four models}
In general for any stochastic process with resetting, it was shown in~\cite{PRL_Reuveni_2016First}, that the MFPT may be expressed in terms of the Laplace transform  $\tilde {F}(s)$ of the distribution of the first passage time (FPT):
\be
\langle T_r \rangle =\frac{1-\tilde {F}(r)}{r\tilde {F}(r)}
\ee
When one is close to ORRVT the optimal resetting rate $r_*$ may be assumed to be small --- this is certainly true for continuous critical and tri-critical transitions and may be a reasonable approximation for discontinuous ones. In such a case, expanding mean $\langle T_r \rangle$ in a Taylor series in $r$ \cite{PRE_Saeed_2019,PRR_Pal_Parsad_Landau_2019} is a good starting point:
\be
\langle T_r \rangle= a_0+a_1r+a_2r^2+a_3r^3+\mathcal{O}(r^4)
\label{eq:small_r_exp}
\ee
This is reminiscent of Landau theory for free energy as a function of the order parameter \cite{Kardar_field}.  The coefficients in Eq.(\ref{eq:small_r_exp}) were derived \cite{PRE_Saeed_2019,PRR_Pal_Parsad_Landau_2019} in terms of the moments of the FPT {\it without resetting}:  $a_0=\langle T \rangle $, $a_1=\langle T \rangle^2-\frac{\langle T^2 \rangle}{2}$, $a_2=\frac{1}{6} \langle T^3 \rangle +\langle T \rangle^3-\langle T \rangle\langle T^2 \rangle$ and $a_3=-\frac{\langle T^4 \rangle}{4!} +\frac{\langle T^3 \rangle\langle T \rangle}{3}+\frac{\langle T^2 \rangle^2}{4}-\frac{3\langle T^2 \rangle\langle T \rangle}{2}+\langle T \rangle^4$. In this limiting expansion  (Eq.(\ref{eq:small_r_exp})), $a_0$ and $a_3$ must be positive to ensure positivity of $\langle T_r \rangle$ at both vanishing $r$ and large $r$ respectively. The sign of the remaining coefficients $a_1$ and $a_2$ decide the type of the ORRVT as we discuss below; although such a discussion may be found in~\cite{PRR_Pal_Parsad_Landau_2019} we provide the following for the ease reading. 

If $a_1$ or $a_2$ or  both are negative $\langle T_r \rangle$ will be minimum at an optimal rate $r=r_*$ which may be located by setting $\frac{\partial \langle T_r\rangle}{\partial r}|_{r=r_*}=a_1+2a_2r_*+3a_3r^2_*=0$. This leads to 
\be
r_{*}=\frac{a_2}{3a_3}\bigg[\sqrt{1-\frac{3a_1a_3}{a^2_2}}\pm1\bigg]
\label{gereral_ORR}
\ee

When $a_1\leq 0$ and $a_2>0$ there is a possibility of a CT by changing a control parameter $\lambda$ (which may be $u_0$ or $v_2$ or $c$ in the different models defined above) at a critical point $\lambda=\lambda_c$.  By setting $a_1=0$ the following important condition is obtained which is very useful to locate the critical point analytically:
\begin{flalign}
  \text{At} \hskip0.01\textwidth \lambda_c:\hskip0.02\textwidth \langle T^2\rangle=2\langle T\rangle^2 \hskip0.01\textwidth\text{and}\hskip0.02\textwidth\langle T^3\rangle>6\langle T\rangle^3
\label{Conti_Trans}
\end{flalign}

Near the transition as $a_1\to 0^-$, we may write $a_1(\lambda)=a_1(\lambda_c)+a_1^{'}(\lambda_c)(|\lambda_c-\lambda|)+\mathcal{O}(\lambda_c-\lambda)^2$ with $a_1(\lambda_c)=0$. This implies $r_{*}\approx |a_1|/2a_2$ (from Eq.\ref{gereral_ORR} considering the root with negative sign for which $r_*\to 0$), and we have $r_*\approx|\lambda_{c}-\lambda|^{\beta}$ with $\beta=1$. Note that this exponent with which ORR vanihes is in contrast to $\beta=1/2$ in the mean field Landau theory of magnetic phase transition~\cite{Kardar_field}.

A special situation arises when $a_1\to 0^-$ and $a_2\to 0^+$. In such a case, the quantity $a_1/a^2_2\to \infty$ which implies $r_*\approx\sqrt{|a_1|/3a_3}$ (from Eq.\ref{gereral_ORR} for the root with the negative sign). Again assuming $a_1\propto|\lambda_c-\lambda|$ we have $r_*\approx|\lambda_{c}-\lambda|^{\beta}$ with $\beta=1/2$. This is again in contrast to $\beta=1/4$ for a TCP in Landau theory of phase transition \cite{Kardar_field}. At this TCP, $\lambda=\lambda_{tc}$, the conditions $a_1=0$ and $a_2 = 0$ gives:
\begin{flalign}
  \text{At} \hskip0.01\textwidth \lambda_{tc}:\hskip0.02\textwidth \langle T^2\rangle=2\langle T\rangle^2 \hskip0.01\textwidth\text{and}\hskip0.02\textwidth\langle T^3\rangle=6\langle T\rangle^3
\label{Trans_TCP}
\end{flalign}

The possibility of DT in ORR arises when $a_1>0$ and $a_2<0$. For such a transition at $\lambda=\lambda_f$ one has a discontinuous jump of $r_*$ from a value $r_f>0$ for $\lambda_f^-$ to $r_*=0$ for $\lambda_f^+$. The DT happens when $\langle T_r\rangle|_{r=0}=\langle T_r\rangle|_{r=r_f}$,   and   $\frac{\partial \langle T_r \rangle}{\partial r}|_{r=r_f}=0$.  If the discontinuous jump in the ``order parameter'' $r_*$ at ORRVT is small, using the Landau expansion (Eq. \ref{eq:small_r_exp}), these conditions lead to $a_1+a_2r_f+a_3r^2_f=0$ and $a_1+2a_2r_f+3a_3r^2_f=0$. Solving the latter equations we get:
\begin{flalign}
 \text{At} \hskip0.01\textwidth \lambda_{f}: \hskip0.02\textwidth  r_{f}=-\frac{a_2}{2a_3}\hskip0.02\textwidth\text{and}\hskip0.02\textwidth a_1=\frac{a_2^2}{4a_3}\hskip0.045\textwidth
\label{Trans_FOT}
\end{flalign}
Using the above Eq.(\ref{Trans_FOT}), one may obtain the jump in $r_*=r_f$ as well as the location of the transition point $\lambda_f$, but only within the small jump approximation. Thus Eq.(\ref{Trans_FOT}) gives approximate values and not exact ones  as Eqn.(\ref{Conti_Trans}) and (\ref{Trans_TCP}). For accurate determination of the DT point and order parameter jump when it is large, studying $\langle T_r\rangle$ (exact or numerical) is better.

To study the possible transitions completely we need to know the moments of the FPT analytically exactly or at least numerically. The moments of the first passage probability distribution $F(x_0,t)$ are also related to the survival probability $Q(x_0,t)=\int^{\infty}_t F(x_0,t) dt$ for the process to survive till time $t$ starting from $x_0$, as follows:
\begin{align}
\langle T^n_r\rangle &=\int^{\infty}_0 t^n F(x_0,t) dt =n\int^{\infty}_0 t^{n-1}Q(x_0,t)dt  \nonumber \\ 
& \equiv n(-1)^{n-1}\frac{\partial^{n-1} q(x_0,s)}{\partial s^{n-1}}\bigg|_{s\to0}
\label{eqn:nth_mements}
\end{align} 
where the second equality comes from integrating by parts and further demanding that $Q(x,t)$ must decay faster than any power law in the limit $t\to\infty$. Thus the knowledge of $\langle T^n_r\rangle$ comes from the knowledge of $Q(x_0,t)$. Here $q(x_0,s)$ is the Laplace transform of $Q(x_0,t)$ with respect to $t$, i.e $q(x_0,s)=\int^{\infty}_{0}\mathrm{d}t Q(x_0,t)e^{-st}$. Note that the $n^{th}$ moment without resetting $\langle T^n\rangle$ can be obtained by setting $r=0$ in the expression of Eq.(\ref{eqn:nth_mements}). The survival probability $Q(x,t)$ for a general initial position $x$ satisfies the backward differential Chapman-Kolmogorov equation (which is the counterpart of the forward Eq.(\ref{eq:Farward_FP})):
\be
\frac{\partial Q}{\partial t}=
D\frac{\partial^{2} Q}{\partial x^{2}}- V^{\prime}(x) \frac{\partial Q}{\partial x}-rQ+rQ_{0},  
\label{eq:Backward_FP}
\ee
Here $Q \equiv Q(x,t)$, $Q_0 \equiv Q(x_0,t)$, and the resetting position $x=x_0$ has been kept distinct from the initial position $x$. On finding $Q(x,t)$, one may replace $x$ by $x_0$ (the particular specified initial position) and solve for $Q(x_0,t)$. That further leads us to the moments through Eq.(\ref{eqn:nth_mements}). The differential equation (\ref{eq:Backward_FP}) is to be analytically or numerically solved for the given initial and boundary conditions. The initial condition is $Q(x,0)=1$. For an AB or a RB at $x=x_{b}$ the conditions are  $Q(x_b,t)=0$ and $Q^{'}(x_b,t)=0$ respectively. 

In the following, we will discuss the methods to obtain $q(x_0,s)$ using Eq.(\ref{eq:Backward_FP}) for the models defined above. For the tent-potential (Eq.(\ref{Tent_pot})), the Eq.(\ref{eq:Backward_FP}) is exactly solvable, but piecewise. However, for the more non-trivial Landau potential (Eq.(\ref{Landau_phi_pot})), we have to take resort to a numerical technique. Both are discussed below in Secs.[\ref{subsec21}] and [\ref{numerical_methos}].
\subsection{\label{subsec21}Analytical solution of the survival probability for the Tent-Potential in Model-I and Model-III}
The potential in Eq.(\ref{Tent_pot}) (as shown in Figs.(\ref{fig:AB_RB_tent_pot}, \ref{fig:AB_AB_tent_pot})) is continuous with piecewise slopes $v_1$ and $-v_2$ on the two side of peak at $x=x_m$. As a result $V^{'}(x)$ is discontinuous at $x=x_m$. The survival probability $Q(x,t)$ may be solved separately on the two sides of the peak such that
\be
\mathrm{Q}(x,t) =
\begin{cases}
 \hfill  Q_-(x,t) ,&  \text{$0\leq x< x_m$},\\
 \hfill Q_+(x,t)  & \text{$L\geq x> x_m$}.
\end{cases}
\label{Full_Survivle_p}
\ee

For $x<x_{m}$, the Eq.(\ref{eq:Backward_FP}) reads

  \begin{align}
    \frac{\partial Q_-(x,t)}{\partial t}=&D\frac{\partial^2 Q_-(x,t)}{\partial x^2}-v_1 \frac{\partial Q_-(x,t)}{\partial x}\nonumber\\
    &-rQ_-(x,t)+rQ_{}(x_{0},t)
\label{eq_BFPE_1}
\end{align}
Similarly for $x>x_{m}$, the (Eq.\ref{eq:Backward_FP}) reads
\begin{align}
  \frac{\partial Q_+(x,t)}{\partial t}=&D\frac{\partial^2 Q_+(x,t)}{\partial x^2}+v_2 \frac{\partial Q_+(x,t)}{\partial x}\nonumber\\
  &-rQ_+(x,t)+rQ_{}(x_{0},t)
\label{eq_BFPE_2}
\end{align}
The two solutions $Q_+$ and $Q_-$ are smoothly connected through the matching conditions $Q_-(x^{-}_m,t)=Q_+(x^{+}_m,t)$ and $Q^{'}_-(x^{-}_m,t)=Q^{'}_+(x^{+}_{m},t)$ at $x=x_m$ (see App.[\ref{App_sec1}] for detailed derivations of the matching conditions).

In the Laplace space the $q_{\pm}(x,s)=\int^{\infty}_{0}e^{-st} Q_{\pm}(x, t)dt$, and the Eqn.(\ref{eq_BFPE_1}) and (\ref{eq_BFPE_2}) lead to the following (using the initial condition $Q_{-}(x,0)=1$ or $Q_{+}(x,0)=1$)
\begin{align}
  D\frac{\partial^2 q_-(x,s)}{\partial x^2}-v_1 \frac{\partial q_-(x,s)}{\partial x}&-(r+s)q_-(x,s)\nonumber\\
  &+rq_{}(x_{0},s)=-1
\label{eq_lap_BFPE_1}
\end{align}
\begin{align}
  D\frac{\partial^2 q_+(x,s)}{\partial x^2}+v_2 \frac{\partial q_+(x,s)}{\partial x}&-(r+s)q_+(x,s)\nonumber\\
  &+rq_{}(x_{0},s)=-1
\label{eq_lap_BFPE_2}
\end{align}
The general solution of these two equations are:
\be
q_{-}(x,s)=A_{1}e^{\alpha^{+}x}+B_{1}e^{\alpha^{-}x}+\frac{1+rq_(x_{0},s)}{r+s}
\label{eq:eq_q-}
\ee
and,
\begin{equation}
q_{+}(x,s)=A_{2}e^{\beta^{+}x}+B_{2}e^{\beta^{-}x}+\frac{1+rq_(x_{0},s)}{r+s}
\label{eq:eq_q+}
\end{equation}

where, $\alpha^{\pm}=\frac{v_1\pm \sqrt{v_1^2+4D(r+s)}}{2D}$ and  $\beta^{\pm}=\frac{-v_2\pm \sqrt{v_2^2+4D(r+s)}}{2D}$. The four constants ($A_1,A_2,B_1$ and $B_2$) may be evaluated using the boundary conditions at $x=0$ and $x=L$ and the two matching conditions $q_{-}(x^{-}_m,s)=q_{+}(x^{+}_m,s)$ and  $q^{'}_{-}(x^{-}_m,s)=q^{'}_{+}(x^{+}_m,s)$ (see App.[\ref{App_sec1}]). These constants depend on the unknown quantity $q(x_0,s)$. When $x_0 < x_m$ we set $q(x_0,s)=q_{-}(x_0,s)$ and $x=x_0$ (the initial position same as the resetting point) in Eq.(\ref{eq:eq_q-}), and solve for $q_{-}(x_0,s)$.  On the other hand when $x_0 > x_m$ we set $q(x_0,s)=q_{+}(x_0,s)$ and $x=x_0$ in Eq.(\ref{eq:eq_q+}), and solve for $q_{+}(x_0,s)$. Depending on whether $x_0< x_m$ or $x_0>x_m$ we use either $q_{-}(x_0,s)$ or $q_{+}(x_0,s)$ to solve for the desired moment $\langle T_r \rangle $. We would find the use of this analytical value of $\langle T_r \rangle $ in the sections below for Model-I and Model-III.

\subsection{\label{numerical_methos} Numerical  solution of the survival probability for the quartic potential in Model-II and Model-IV}
For the potential $V(x)$ (Eq.\ref{Landau_phi_pot}), also shown in Figs.(\ref{fig:AB_RB_Ising_pot}) and (\ref{fig:AB_AB_Ising_pot})), the equation for $q(x,s)$ is as follows:
\begin{align}
  D\frac{\partial^2 q(x,s)}{\partial x^2}-V^{'}(x)\frac{\partial q(x,s)}{\partial x}&-(r+s)q(x,s)\nonumber\\
  &+rq(x_{0},s)=-1
\label{eq_lap_BFPE_Landau}
\end{align}
It is hard to solve analytically because of the form of $V^{'}(x)=4x^3-2bx-c$. Even solving this numerically is a challenge because $q(x,s)$ may be obtained by the numerical integration of the above equation for an assumed $q(x_0,s)$, which in turn is not known before $q(x,s)$ is solved.

We handle the problem using the following method developed in Ref.\cite{PRE_Saeed_2019}. First we make the Eq.(\ref{eq_lap_BFPE_Landau}) homogeneous by introducing a new function $y(x,s) = q(x,s)-\frac{r q(x_0,s) +1}{(r+s)}$, which looks as:
\be
D\frac{\partial^2 y(x,s)}{\partial x^2}-V^{'}(x)\frac{\partial y(x,s)}{\partial x}-(r+s)y(x,s)=0
\label{eq_lap_y_BFPE_Landau}
\ee

Note that the absorbing boundary (AB) condition $q(x_{\text{AB}},s)=0$ translates to $y(x_{\text{AB}},s)=-\frac{r q(x_0,s) +1}{(r+s)}$ while reflecting boundary (RB) condition on $q^{'}(x_{\text{RB}},s)=0$ implies $y^{'}(x_{\text{RB}},s)=0$. Thus the unknown value $q(x_0,s)$ still lingers around through the AB condition. Next we proceed to define a scaled function $\tilde{y}(x,s)=y(x,s)/y(x_{\text{AB}},s)$. The differential equation for $\tilde{y}(x,s)$ is: 
\be
\frac{\mathrm{d}^{2} \tilde{y}}{\mathrm{d} x^{2}}=\frac{V^{'}(x)}{D}\frac{\mathrm{d} \tilde{y}}{\mathrm{d} x}+\frac{(r + s)}{D} \tilde{y}
\label{eq:NUmerical_Sol}
\ee
It is remarkable that the AB condition $\tilde{y}(x_{\text{AB}},s)=1$ and hence the solution of Eq.(\ref{eq:NUmerical_Sol}) becomes free of the unknown input value $q(x_0,s)$. The RB condition is $\tilde{y}^{'}(x_{\text{RB}},s)=0$. If one may find $\tilde{y}(x,s)$ numerically using $\tilde{y}(x_0,s)$ one may obtain $q(x_0,s)$ from the following formula
\begin{flalign}
  q(x_0,s) &= (\tilde{y}(x_0,s) - 1) y(x_{AB},s)\nonumber\\
  &=-(\tilde{y}(x_0,s) - 1)\frac{r q(x_0,s) +1}{(r+s)},\nonumber\\
    \implies q(x_0,s) &= \frac{1-\tilde{y}(x_0, s)}{s+r\tilde{y}(x_0,s)}.
    \label{eq:NUmerical_Sol_s0}
\end{flalign}
But finding $\tilde{y}(x,s)$ for all possible values of $s$ is a tedious job and in fact we do not need that. What we need is  $\langle T_r \rangle=q(x_0,s)|_{s=0}$ and $q(x_0,0)$ is related to $\tilde{y}(x_0,0)$. 
After setting $s=0$, we solve the Eq.~(\ref{eq:NUmerical_Sol}) using {\it NDSolve} technique in {\it Mathematica} which includes {\it ExplicitRungeKutta} method to obtain $\tilde{y}(x,0)$.  Further using the relation below (which follows from Eq.(\ref{eq:NUmerical_Sol_s0})) the mean time with resetting can be found from the numerical value of $\tilde{y}(x_0,0)$:
\be
\langle T_r \rangle=q(x_0,0) = \frac{1-\tilde{y}(x_0, 0)}{r\tilde{y}(x_0,0)}.
\label{MFPT_N_V}
\ee
The mean time $\langle T_r \rangle$ can then be studied as a function of $r$, for different parameter values $b$ and $c$ (Eq.(\ref{Landau_phi_pot})), to obtain the continuous and discontinuous ORRVT, in Model-II and Model-IV.

\section{Continuous transitions in the models with one absorbing boundary and one reflecting boundary:}
\subsection{\label{Model1_discus}Model-I}
To locate the continuous ORRVT in this model for any given values of $v_1$ and $v_2$ we need to obtain $\langle T\rangle$ and $\langle T^2 \rangle$ in the absence of resetting and substitute in the Eq.(\ref{Conti_Trans}). As we discussed in the Sec.[\ref{subsec21}] for the discontinuous $V^{'}(x)$ we have piecewise solutions of $q_{+}(x,s)$ and $q_{-}(x,s)$. Consequently the moments of FPT will also have piecewise solutions. Noting that $\langle T\rangle_{\pm}=q_{\pm}(x,0)|_{r=0}$ and $\langle T^2\rangle_{\pm}=-2\frac{\partial q_{\pm}(x,s)}{\partial s}\big|_{s\to 0, r=0}$, from Eqn. (\ref{eq_lap_BFPE_1}) and (\ref{eq_lap_BFPE_2}) we get the following:

\be
\mathrm{For}\hskip0.02\textwidth x<x_m: \hskip0.0\textwidth 
\begin{cases}
 D\frac{d^2  \langle T \rangle_{-}}{d x^2}-v_1\frac{d \langle T \rangle_{-}}{d x}=-1,\\[5pt]
D\frac{d^2  \langle T^2 \rangle_{-}}{d x^2}-v_1\frac{d \langle T^2 \rangle_{-}}{d x}=-2\langle T \rangle_{-}.
\end{cases}
\label{diff_tent_Mom_l}
\ee

\be
\mathrm{For}\hskip0.02\textwidth x>x_m: \hskip0.0\textwidth 
\begin{cases}
 D\frac{d^2  \langle T \rangle_{+}}{d x^2}+v_2\frac{d \langle T \rangle_{+}}{d x}=-1,\\[5pt]
D\frac{d^2  \langle T^2 \rangle_{+}}{d x^2}+v_2\frac{d \langle T^2 \rangle_{+}}{d x}=-2\langle T \rangle_{+}.
\end{cases}
\label{diff_tent_Mom_R}
\ee

The AB is situated at $x=0$ (i.e. $u=0$) and the RB is at $x=L$ (i.e. $u=1$). At the AB, the moments satisfy the conditions $\langle T \rangle_{-}|_{x=0}=0$, and $\langle T^2 \rangle_{-}|_{x=0}=0$, while at RB, the conditions are $\langle T \rangle^{'}_{+}|_{x=L}=0$, and $\langle T^2 \rangle^{'}_{+}|_{x=L}=0$. Across $x=x_m$ the matching conditions are $\langle T \rangle_{-}|_{x=x_m}=\langle T \rangle_{+}|_{x=x_m}$, $\langle T\rangle^{'}_{-}|_{x=x_m}=\langle T \rangle^{'}_{+}|_{x=x_m}$, $\langle T^2 \rangle_{-}|_{x=x_m}=\langle T^2 \rangle_{+}|_{x=x_m}$ and $\langle T^{2} \rangle^{'}_{-}|_{x=x_m}=\langle T^{2} \rangle^{'}_{+}|_{x=x_m}$ (which follow from the discussions in App.[\ref{App_sec1}]). After applying these conditions the desired solutions of Eqn.(\ref{diff_tent_Mom_l}) and (\ref{diff_tent_Mom_R}) are shown in the App.[\ref{App_sec2_M1}]. 

\begin{figure}
  \begin{minipage}{.5\textwidth}
\begin{subfigure}{0.45\textwidth}
     \vspace*{0.1cm}
    \hspace*{-0.5cm}
    \vspace*{0.1cm}
    \includegraphics[width = 1.1\textwidth,height=0.12\textheight]{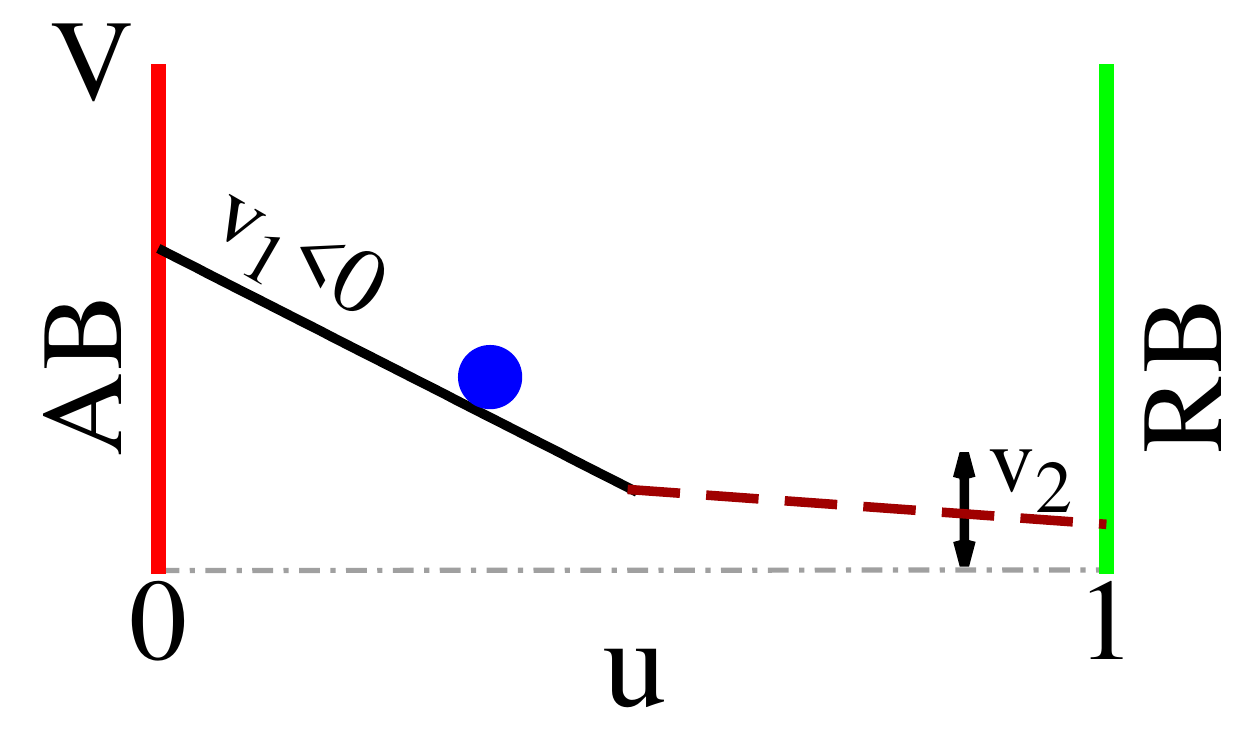}
    \caption[]{} 
    \label{fig:Model1_pot1}
   \end{subfigure}
\begin{subfigure}{0.45\textwidth}
    \includegraphics[width = 1.1\textwidth,height=0.13\textheight]{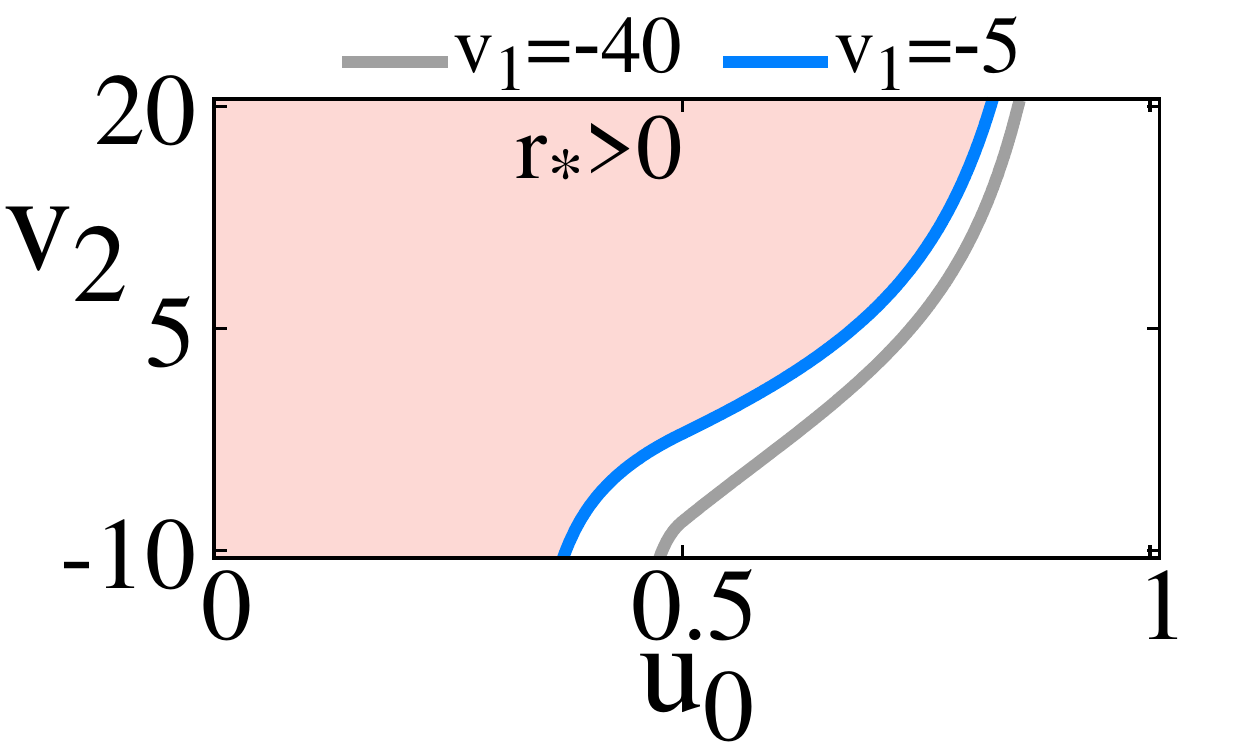}
    \caption[]{} 
    \label{fig:Model1_trans1}
  \end{subfigure}
\begin{subfigure}{0.45\textwidth}
     \vspace*{0.1cm}
    \hspace*{-0.5cm}
    \vspace*{0.1cm}
    \includegraphics[width = 1.1\textwidth,height=0.12\textheight]{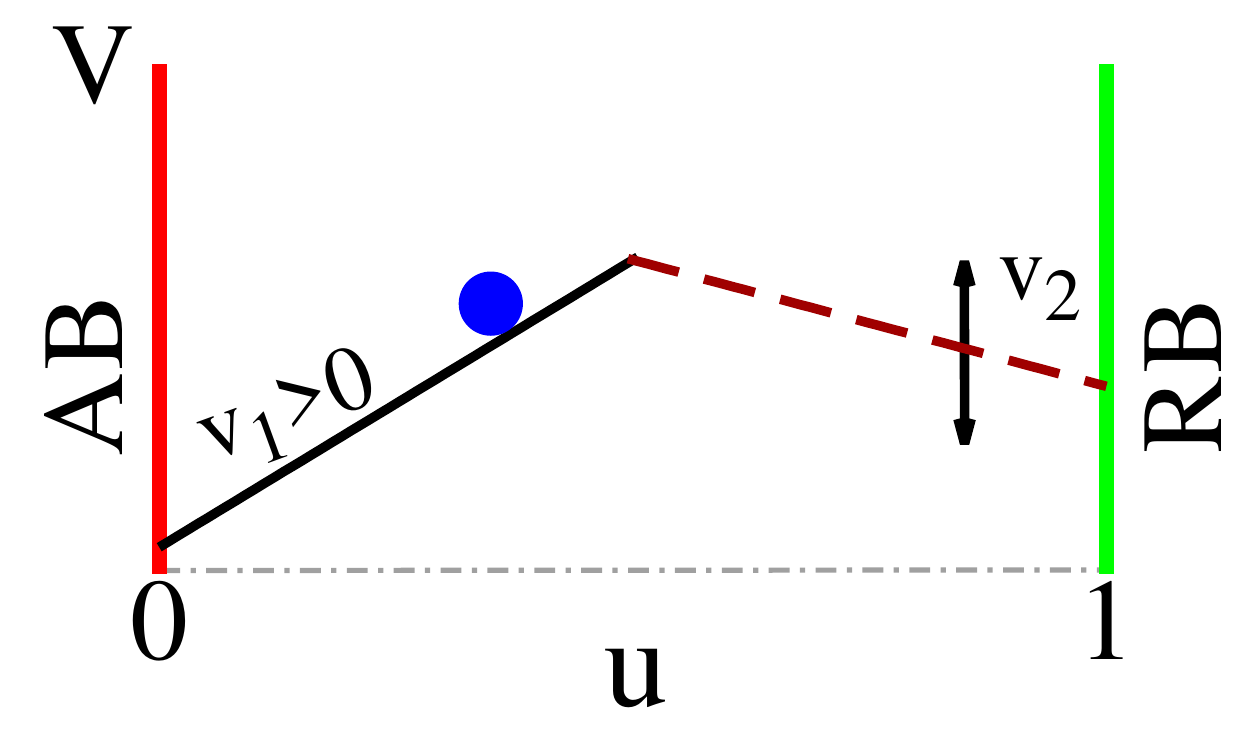}
    \caption[]{} 
    \label{fig:Model1_pot2}
   \end{subfigure}
\begin{subfigure}{0.45\textwidth}
    \includegraphics[width = 1.1\textwidth,height=0.13\textheight]{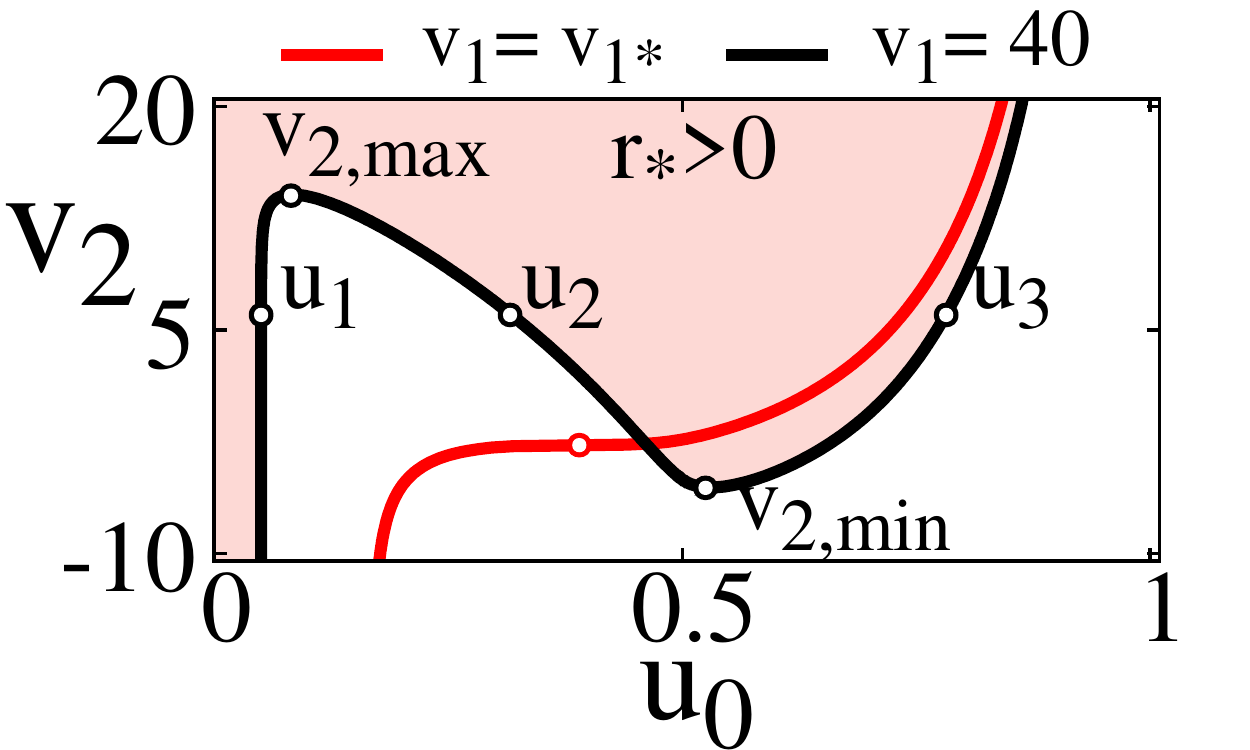}
    \caption[]{} 
    \label{fig:Model1_trans2}
  \end{subfigure}
\caption{We show the tent potential as a function of the scaled distance $u$ when there is no peak (in (a)) and when the peak is sharp (in (c)). In (b) and in (d), we show the critical lines in the space of $v_2 - u_0$ corresponding to the scenarios in (a) and (c) respectively. In (b) for $v_1=-5$ the region with $r_*>0$ is shaded in purple. Similarly, in (d) for $v_1=40$ the region with beneficial resetting strategy is shown in purple. For the figures we used $D=1$, $x_m=0.5$ and $L=1$. For $v_1>v^*_1=11.347$ the possibility of multiple critical points arise, as a function of $u_0$ for a given $v_2$. For $v_1=40$ in (d), $u_{1}=0.0506$ , $u_{2}=0.2291$ , $u_{3}=0.8124$ and $v_{2,min}=-5.60$ , $v_{2,max}=14.20$.
  }
    \label{fig:Model1_tran_line}
\end{minipage}
  \hfill
  \begin{minipage}{0.4\textwidth}
  \end{minipage}
\end{figure}

When $x_0<x_m$, the Eq.(\ref{Conti_Trans}) gets modified to $\langle T^2\rangle_{-}=2\langle T\rangle^{2}_{-} $, and when  $x_0>x_m$, we have $\langle T^2\rangle_{+}=2\langle T\rangle^{2}_{+}$. These conditions on the moments help us obtain the exact locations ($2-d$ surfaces) of the continuous ORRVT in the $3-d$ parameter space of ($v_1, v_2, u_0$). We show projected transition lines for fixed values of $v_1$ in the $v_2 - u_0$ plane in Fig.(\ref{fig:Model1_trans1}) and Fig.(\ref{fig:Model1_trans2}). We systematically vary $v_1$ from negative values (Fig.(\ref{fig:Model1_pot1})), to high positive values (Fig.(\ref{fig:Model1_pot2})). In the scenario of Fig.(\ref{fig:Model1_pot1}), we see (in Fig.(\ref{fig:Model1_trans1})) that for any $v_2$ (positive  or negative) on varying the scaled initial position $u_0$ we have a single CT from the beneficial resetting phase (with $r_*>0$) to the unbeneficial phase ($r_*=0$) at some critical point $u_{0c}=u_{0c}(v_1,v_2)$.

On the other hand for the scenario in Fig.(\ref{fig:Model1_pot2}) for large positive $v_1$ above a certain $v^{*}_1$ we have a new possibility.  Within a certain range of $v_2\in(v_{2,min},v_{2,max})$, for a constant $v_2$, we have multiple CTs as we vary $u_0$. First $r_*>0$ changes to $r_*=0$ phase at $u_{0c}=u_{1}$, then there is a re-entrance to the $r_*>0$ phase at $u_{0c}=u_{2}$, and finally there is a transition to the $r_*=0$ phase at $u_{0c}=u_{3}$. Thus for the case of a deep valley near the AB ($u_0=0$), corresponding to large $v_1$, the advantage of resetting is crucially determined by the location of the resetting point $u_0$. If one is near the AB or near the hill top it seems resetting is beneficial, but otherwise not.

While for very large $v_1$ and $v_2$, since the peak at $x_m$ is sharp, it is expected that resetting would help in crossing the barrier and attaining first passage irrespective of the location $u_0$ --- hence we see $r_*>0$ for most values of $u_0$. Likewise, for small values of $v_2$, the peak is not so sharp, and so resetting is mostly unnecessary --- as a result for most values of $u_0$ we have $r_*=0$. Yet the existence of multiple regions where resetting is either beneficial or not,  for $v_2\in(v_{2,min},v_{2,max})$, is rather interesting and not intuitively expected {\it a prori}. 

Is a DT possible in this model? To check this we derived the exact expressions of MFPT in the presence of resetting (see App.[\ref{App_sec2_M1}]) using Eqs. (\ref{eq_lap_BFPE_1}) and (\ref{eq_lap_BFPE_2}) and appropriate boundary and matching conditions --- note that $\langle T_r \rangle_{-}= q_-(x,s)|_{s=0}$ and $\langle T_r \rangle_{+}= q_+(x,s)|_{s=0}$. Plotting the MFPT as a function of $r$ for finite values of $v_1, v_2, u_0,$ we found that there exists no DT. We also plotted the parameter $a_2$ in Landau theory (Eq.(\ref{eq:small_r_exp})) as a function of $u_0$ between the two extremes $v_2\to \pm\infty$ --- we found that always $a_2>0$, implying the nonexistence of DT in Model-I (see Fig.(\ref{fig:Model1_a2_vs_u0}) in App.[\ref{App_sec2_M1}]).


\subsection{Model-II}
\begin{figure}
    \begin{minipage}{.5\textwidth}
   \begin{subfigure}{0.45\textwidth}
    \vspace*{0.0cm}
    \hspace*{-1.0cm}
    \vspace*{0.2cm}
    \includegraphics[width = 1.2\textwidth,height=0.12\textheight]{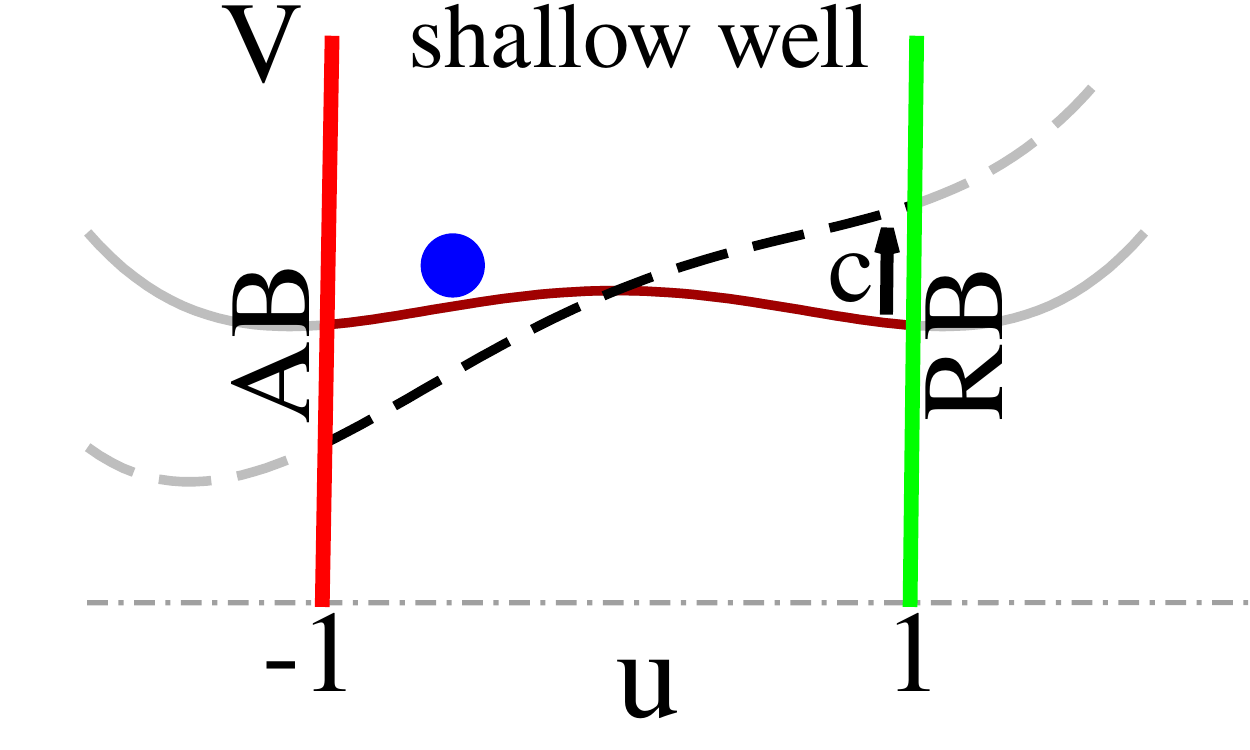}
    \caption[]{} 
    \label{fig:Model2_part1}
   \end{subfigure}
  \begin{subfigure}{0.45\textwidth}
    \includegraphics[width = 1.1\textwidth,height=0.126\textheight]{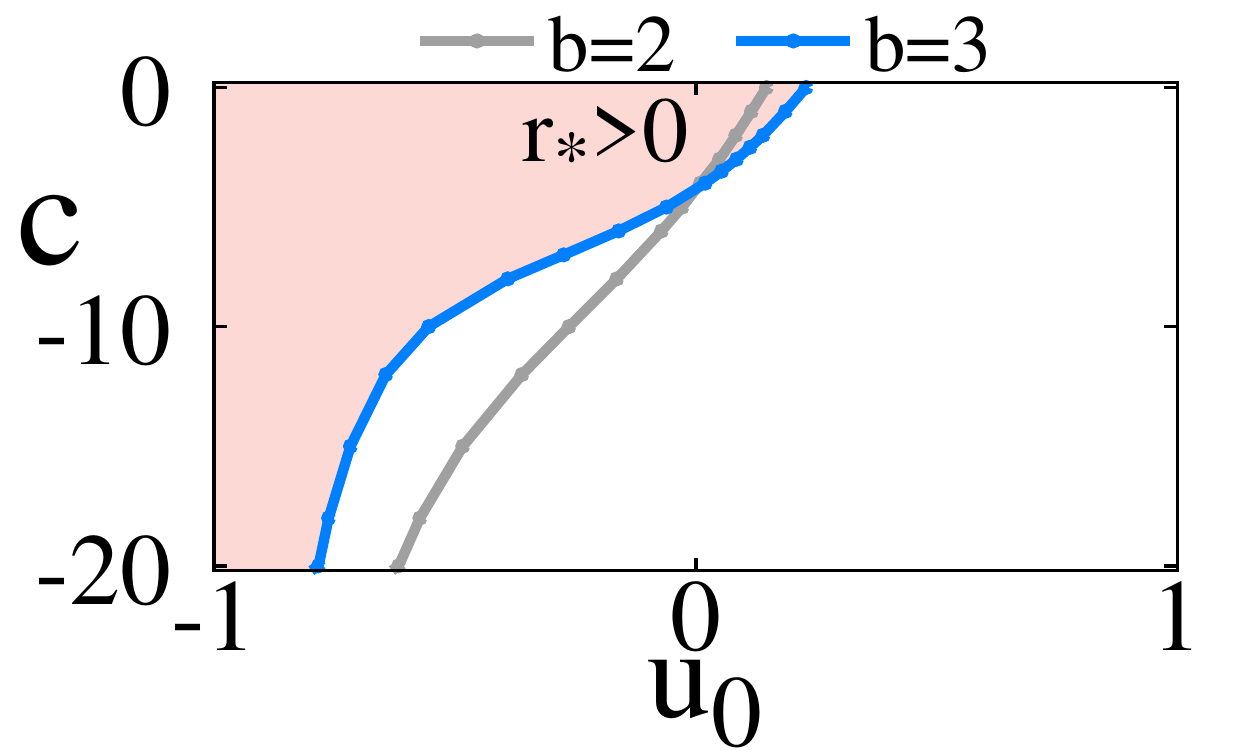}
    \caption[]{} 
    \label{fig:Model2_tran_line1}
  \end{subfigure}
  \begin{subfigure}{0.45\textwidth}
    \vspace*{0.0cm}
    \hspace*{-1.0cm}
    \vspace*{0.2cm}
    \includegraphics[width = 1.2\textwidth,height=0.123\textheight]{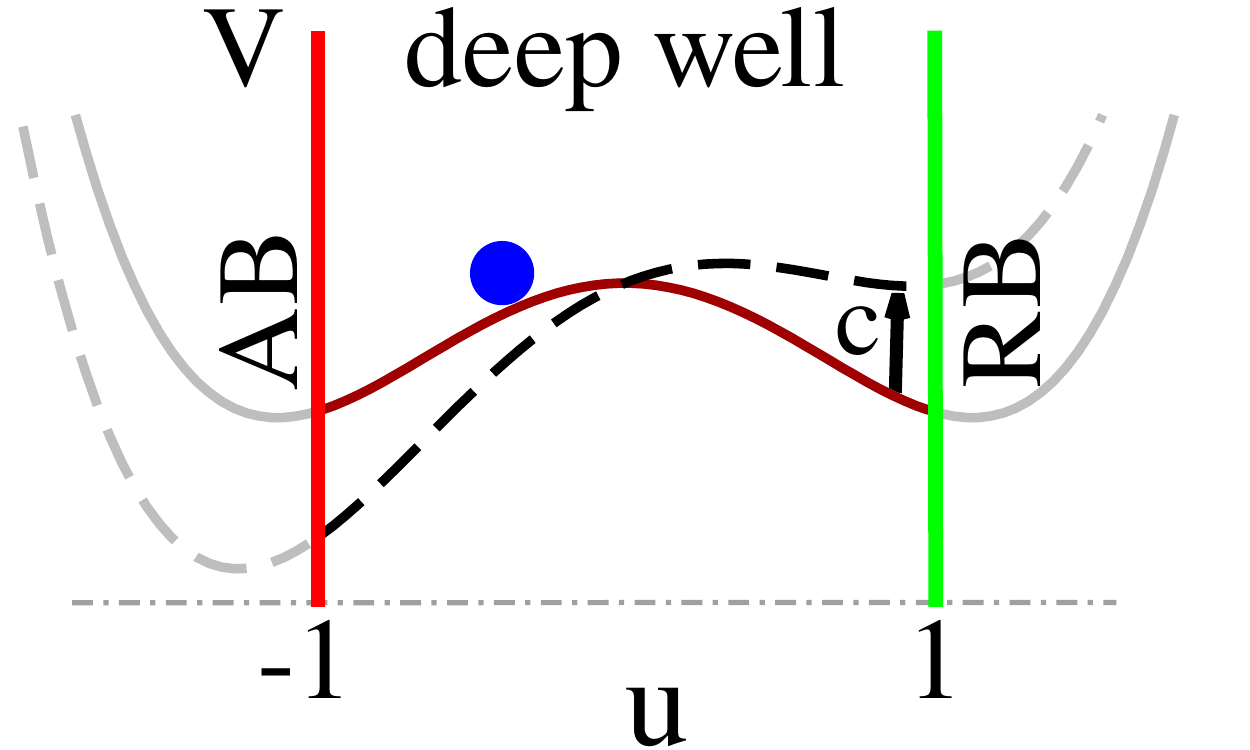}
    \caption[]{} 
    \label{fig:Model2_part2}
   \end{subfigure}
  \begin{subfigure}{0.45\textwidth}
    \includegraphics[width = 1.1\textwidth,height=0.126\textheight]{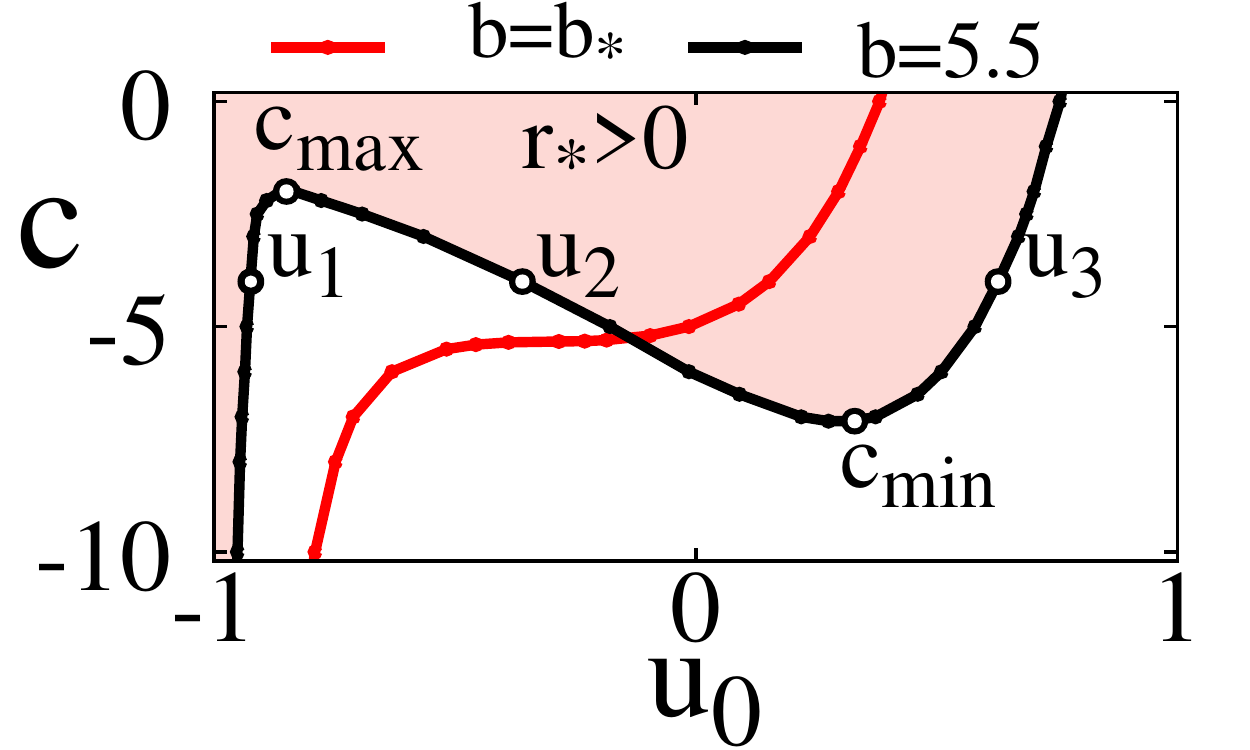}
    \caption[]{} 
    \label{fig:Model2_tran_line2}
  \end{subfigure}
  \caption{We show the quartic potential (Eq.(\ref{Landau_phi_pot})) in the region between AB at $u=-1$ and RB at $u=1$ for a temperature close to $T_c$ (a) and another much below $T_c$ (c). The part of the potential which is outside the allowed region have shown in lighter gray lines. Application of field $|c|>0$ make the potential wells asymmetric as shown by the dashed lines. The critical lines in the $c-u_0$ plane are shown in (b) and (d). The phase with $r_*>0$ is shown in purple color for $b=2$ in (b) and $b=5.5$ in (d). The values of $c_{max}=-2.0$,  $c_{min}=-7.1$ and the three critical points corresponding to $c=4$ are $u_1=-0.92$, $u_2=-0.36$ and $u_3=0.63$. For $b<b_*=3.95$ (shown in red line), there is only a single critical point for a given $c$. The temperature dependent diffusion constant used in this study is $D=6T/T_c$.}
  \label{fig:Model2_study}
  \end{minipage}
  \hfill
  \begin{minipage}{0.4\textwidth}
  \end{minipage}
\end{figure}

Unlike Model-I, the Model-II is not exactly solvable because of the quartic potential $V(x)$ (Eq.(\ref{Landau_phi_pot})), but the MFPT  $\langle T_r \rangle$ can be calculated numerically following the method discussed in Sec[\ref{numerical_methos}]. The boundary conditions for this model are $\langle T_r \rangle=0$ at $u=-1$ and  $\langle T_r \rangle^{'}=0$ at $u=+1$, where $u=\frac{x\sqrt{2}}{\sqrt{b}}$ is the scaled magnetization. Note that we need no matching condition as in Model-I as $V^{'}(x)$ is continuous everywhere. The diffusion constant appearing in the differential equation that we solve (Eq.(\ref{eq:NUmerical_Sol})) is temperature dependent for the equilibrium magnetic system in the absence of resetting --- so we set $D\propto \frac{T}{T_c}=1-\frac{b}{6}$. For any given temperature ($b$), field ($c$), and initial magnetization ($u_0$), the $\langle T_r \rangle$ may be plotted as a function of $r$ to obtain its minimum, at the ORR $r=r_*$. For a given temperature ($b=$constant), this leads to the ORRVT lines in the two-dimensional $c - u_0$ plane as shown in Figs.(\ref{fig:Model2_tran_line1}) and (\ref{fig:Model2_tran_line2}).

In Fig.(\ref{fig:Model2_part1}) we consider a shallow double potential corresponding to temperature closer to $T_c$. For $c=0$ it is symmetric, while for $c>0$ asymmetry develops. Note we have the AB and hence first passage at a location near the negative magnetization minimum. Corresponding to this scenario, we find that for any given field $c\leq 0$, there is a single CT from the $r_*>0$ phase to the $r_*=0$ phase (see Fig.(\ref{fig:Model2_tran_line1})). The RB at $u=1$ helps in the first passage by providing confinement so effectively that resetting strategy is not helpful, unless $u_0$ is close to the AB where resetting helps to curtail long excursion away from the target. 

The above scenario changes as temperature becomes low and consequently, potential wells deepen (see Fig.(\ref{fig:Model2_part2})). For $b>b_*$ the critical transition line in the $c-u_0$ plane has two turning points. As a result for a given magnetic field $c\in(c_{min}, c_{max})$, there are three critical points $u_1,u_2$ and $u_3$. For reset points $u_0\in(u_1,u_2)$ and $u_0\in(u_3,1)$, resetting does not help. On the other hand for $u_0\in(-1,u_1)$ (near AB) and $u_0\in(u_2,u_3)$ (near the potential barrier) the resetting strategy is indeed beneficial and $r_*>0$.

Note that this model represents a realistic magnetic system, and at very low temperature $T\to 0$ (i.e. $b\to 6$) barrier crossing is indeed a challenge to go from a positive magnetization state ($u_0\in(0,1)$) to the negative magnetization state $u_0=-1$. We see three things  in Fig.(\ref{fig:Model2_tran_line2}). At high magnetic field ($c\ll 0$), the asymmetry in the potential can drive first passage efficiently such that resetting is unnecessary for most $u_0$. On the other hand, for very low values of field $c$, resetting is a good strategy to additionally help in barrier crossing. Finally, at intermediate values of $c$, it is most interesting that resetting only helps if done to specific ranges of the initial magnetization.

\section{Discontinuous transitions and multiple tri-critical points in models with two absorbing boundaries:}

\subsection{\label{Model3_discus}Model-III}
The mathematical procedure to study continuous ORRVT for this model with tent potential (piecewise linear $V(x)$ and discontinuous $V^{'}(x)$) has been discussed in Sec.[\ref{Model1_discus}]. To obtain the CT, one needs to solve for $\langle T \rangle_{-}$, $\langle T \rangle_{+}$, $\langle T^2 \rangle_{-}$ and $\langle T^2 \rangle_{+}$ without resetting using Eqn. (\ref{diff_tent_Mom_l}) and (\ref{diff_tent_Mom_R}) with matching conditions of MFPT and its derivative at $x=x_m$. The boundary conditions in this model are different from Model-I. We have $\langle T \rangle_{-}|_{x=0}=0$, $\langle T^2 \rangle_{-}|_{x=0}=0$, $\langle T \rangle_{+}|_{x=L}=0$, and $\langle T^2 \rangle_{+}|_{x=L}=0$, at both $x=0$ (i.e. $u=0$) and $x=L$ (i.e. $u=1$). The exact expressions are shown in App.[\ref{App_M3}]. Setting  $\langle T^2 \rangle_{-}=2\langle T \rangle^2_{-}$ for $x<x_m$ and $\langle T^2 \rangle_{+}=2\langle T \rangle^2_{+}$ for $x>x_m$ we get the surfaces of CT in the $3-d$ $(v_1, v_2, u_0)$ space.

To obtain the locations of the DTs which arise in this model, we solve Eqs. (\ref{eq_lap_BFPE_1}) and (\ref{eq_lap_BFPE_2}) for $\langle T_r \rangle_{-}= q_-(x,s)|_{s=0}$ and $\langle T_r \rangle_{+}= q_+(x,s)|_{s=0}$ with resetting, with suitable boundary and matching conditions (mentioned above). The results may be seen in App.[\ref{App_M3}]. Then we plot MFPT as a function of the resetting rate $r$ and obtain the discontinuous jump in $r_*$, whenever they arise. We would like to point out that the jump in $r_*$ may be quite large, and the analytical formula obtained from small order parameter expansion (Eq.(\ref{Trans_FOT})) may not lead to numerically accurate answers (see Fig.(\ref{fig:Model3_MFPT_vs_r}) in App.[\ref{App_M3}]). Hence the procedure outlined above for DT is better. 
\begin{figure}[!ht]
  \begin{subfigure}{0.23\textwidth}
        \vspace*{0.1cm}
    \hspace*{-0.5cm}
    \vspace*{0.11cm}
\includegraphics[width =1.1\textwidth,height=0.11\textheight]{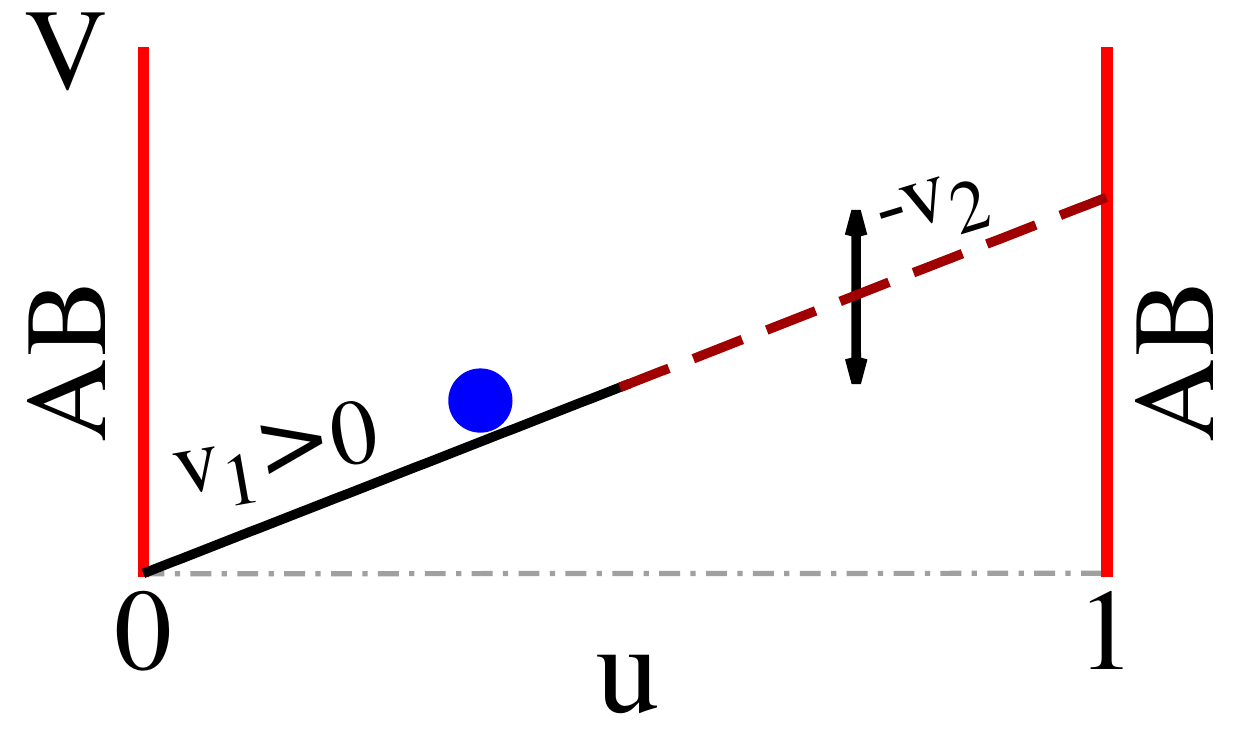}
    \caption[]{} 
    \label{fig:Model3_Pot_line1}
 \end{subfigure}
   \begin{subfigure}{0.23\textwidth}
    \includegraphics[width = 1.1\textwidth,height=0.12\textheight]{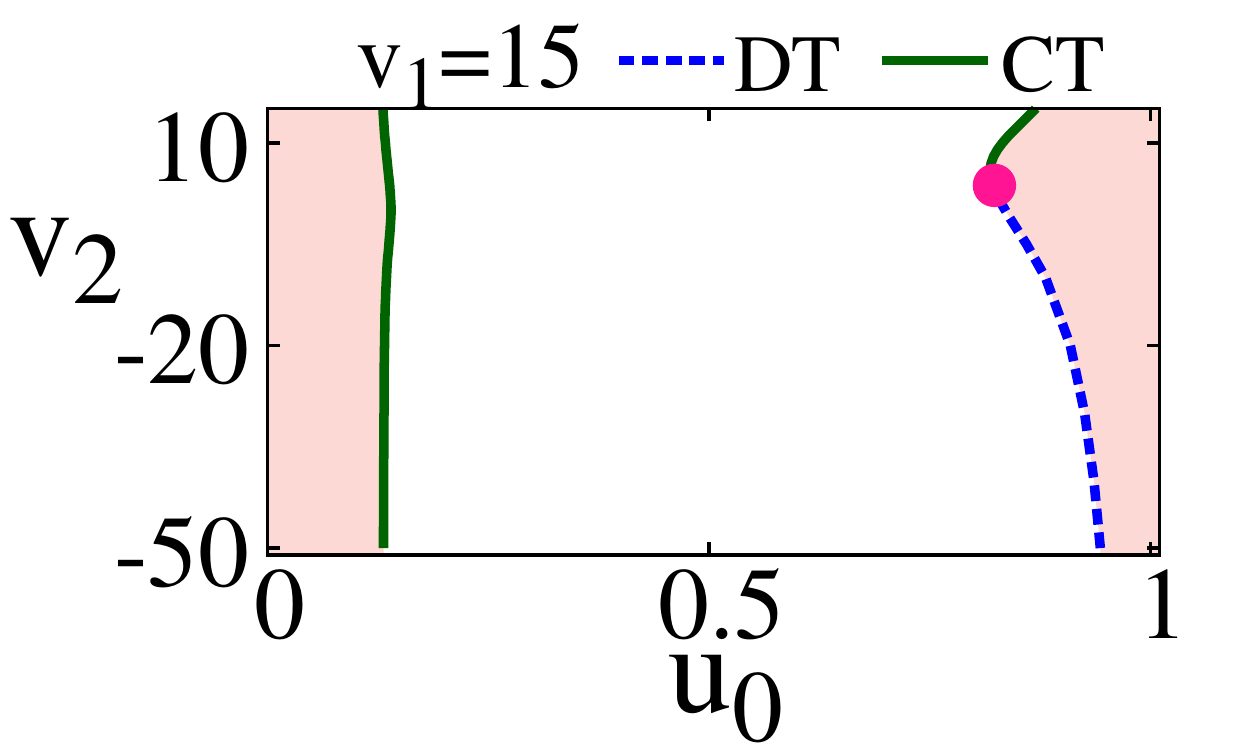}
   \caption[]{} 
    \label{fig:Model3_tran_line1}
  \end{subfigure}
   \caption{(a) The tent potential is shown for $v_2=-v_1$. The scenario remains similar unless $v_2>0$ and large. In (b), the ORRVT lines in the $v_2-u_0$ plane is shown, for $v_1=15$. There is a left branch of CT, and a right branch of DT which meets a continuous line at TCP ($v_{2,\text{TCP}}=4.57$, $u_{0,\text{TCP}}=0.82$). Here we used $D=1$, $x_m=0.5$ and $L=1$.}
    \label{fig:Model3_tran_line_A_Pal}
\end{figure}

If we set $v_2=-v_1$, then the potential looks as in Fig.(\ref{fig:Model3_Pot_line1}) and the problem becomes identical to the one studied in Ref.\cite{PRR_Pal_Parsad_Landau_2019}. In  Fig.(\ref{fig:Model3_tran_line1})  for $v_1=15$ and $v_2<0$ as we vary $u_0$, the ORR goes from a nonzero value to zero continuously at a critical point, and then undergoes a discontinuous jump at a second transition point. The DT line terminates at a TCP (marked in pink). For $v_2>v_{2,\text{TCP}}=4.57$ there are two CTs as a function of $u_0$.

 \begin{figure}
    \begin{minipage}{.5\textwidth}
      \begin{subfigure}{0.45\textwidth}
         \vspace*{0.12cm}
    \hspace*{-0.5cm}
    \vspace*{0.11cm}
   \includegraphics[width = 1.1\textwidth,height=0.11\textheight]{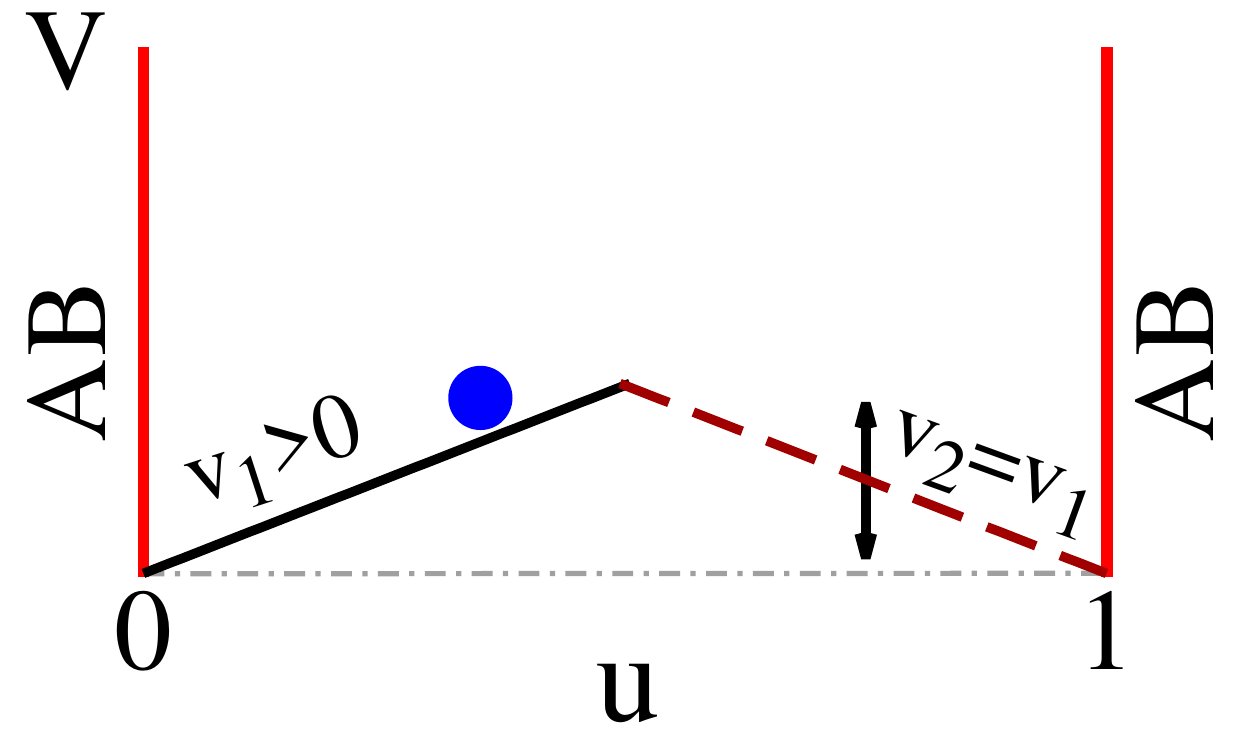}
   \caption[]{} 
    \label{fig:Model3_Pot_line2}
 \end{subfigure}
\begin{subfigure}{0.45\textwidth}
    \includegraphics[width = 1.1\textwidth,height=0.12\textheight]{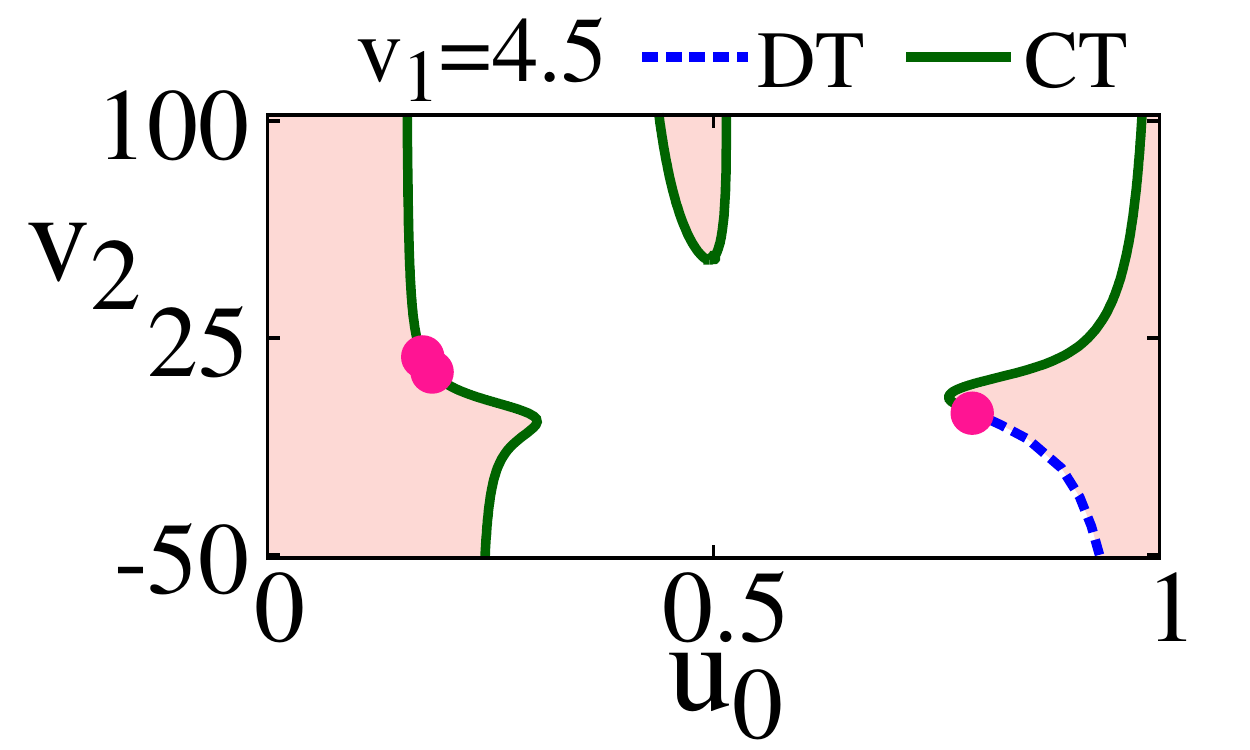}
   \caption[]{} 
    \label{fig:Model3_tran_line2}
   \end{subfigure}
\begin{subfigure}{0.45\textwidth}
   \vspace*{0.12cm}
    \hspace*{-0.5cm}
    \vspace*{0.11cm}
    \includegraphics[width =1.1\textwidth,height=0.11\textheight]{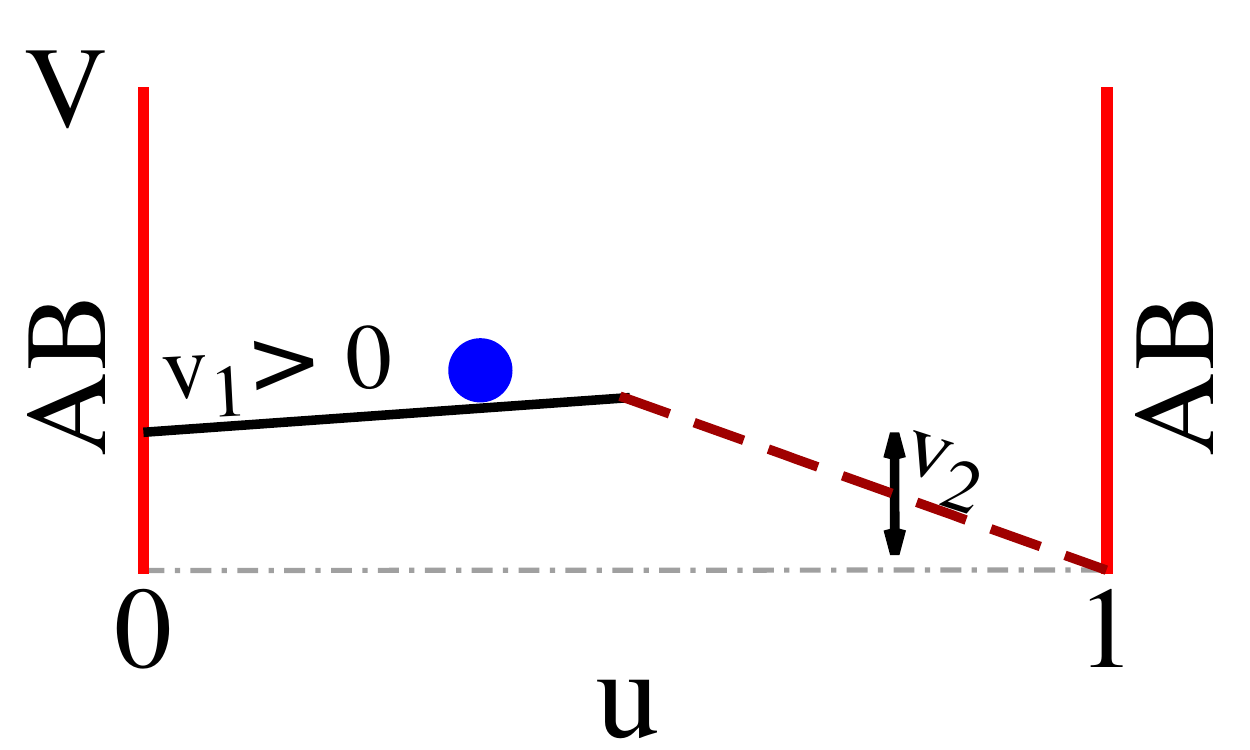}
    \caption[]{} 
    \label{fig:Model3_Pot_line3}
  \end{subfigure}
  \begin{subfigure}{0.45\textwidth}
    \includegraphics[width = 1.1\textwidth,height=0.12\textheight]{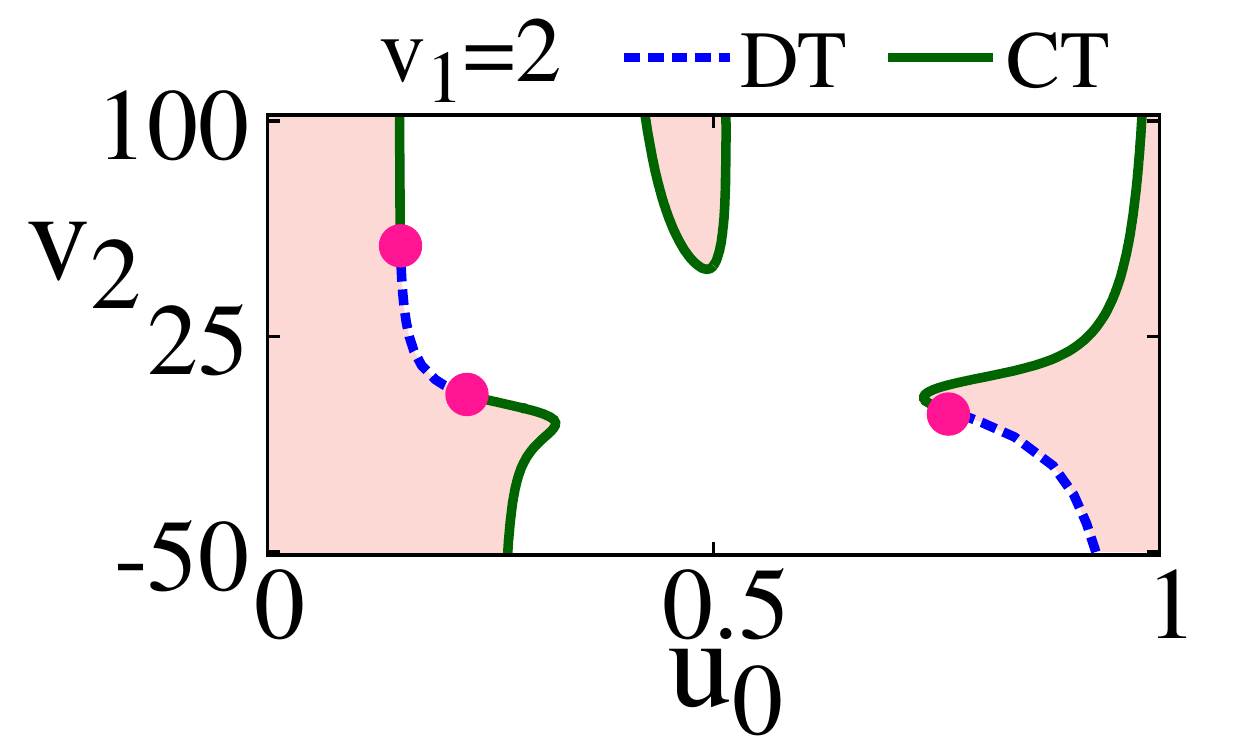}
   \caption[]{} 
    \label{fig:Model3_tran_line3}
   \end{subfigure}
  \begin{subfigure}{0.45\textwidth}
     \vspace*{0.12cm}
    \hspace*{-0.5cm}
    \vspace*{0.11cm}
   \includegraphics[width = 1.1\textwidth,height=0.11\textheight]{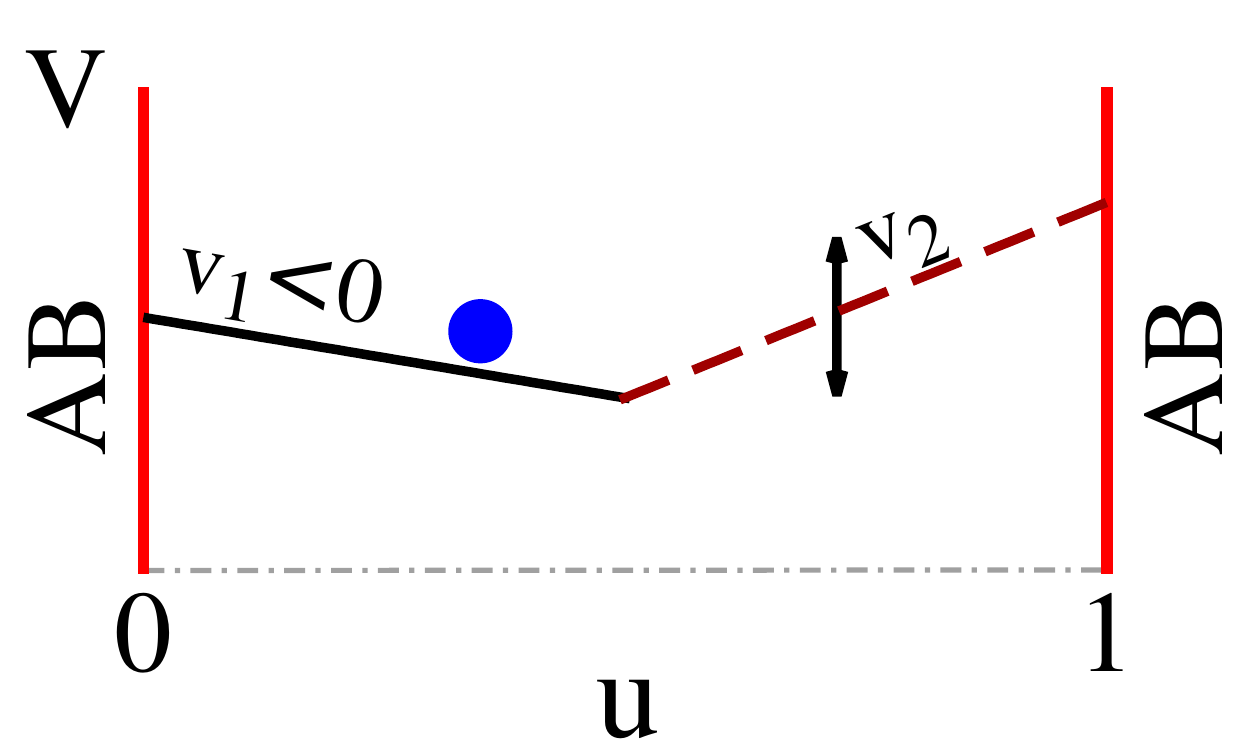}
   \caption[]{} 
    \label{fig:Model3_Pot_line4}
 \end{subfigure}
 \begin{subfigure}{0.45\textwidth}
    \includegraphics[width = 1.1\textwidth,height=0.12\textheight]{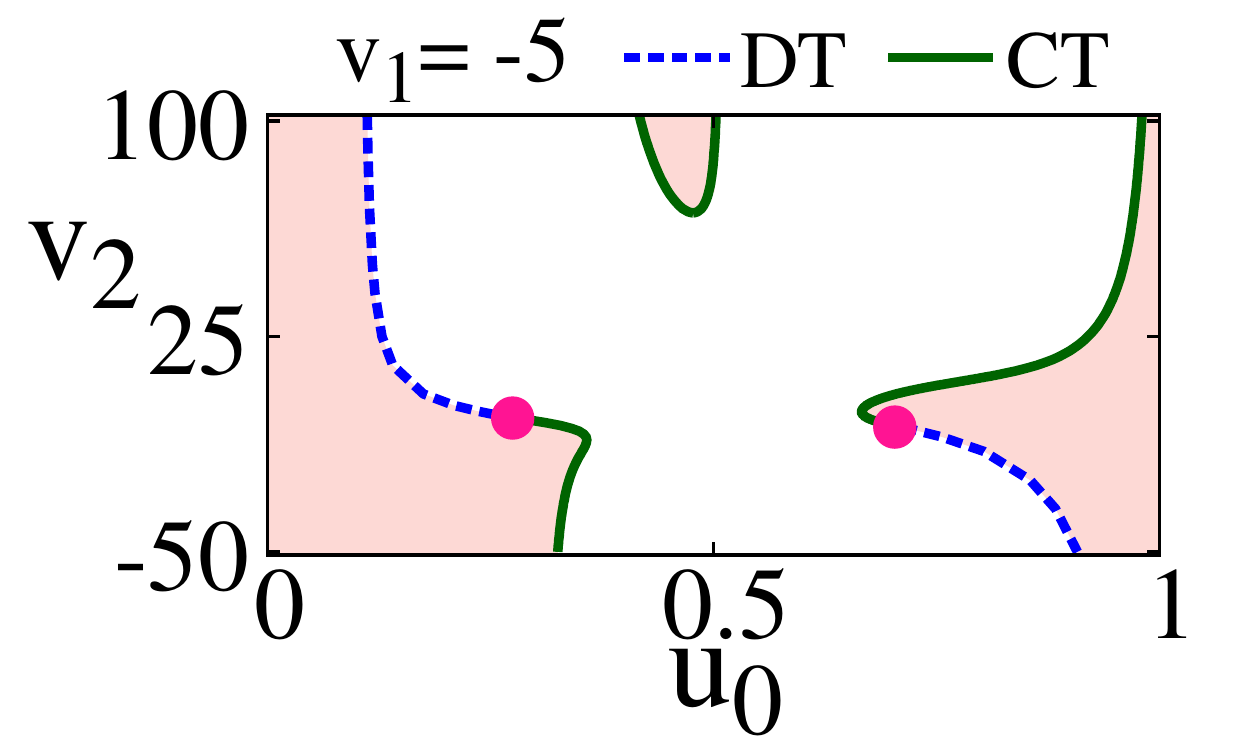}
    \caption[]{} 
    \label{fig:Model3_tran_line4}
   \end{subfigure}
   \caption{We make the slope $v_1$ go systematically from positive to negative values as shown in (a), (b) and (c). The corresponding ORRVT lines in the $v_2-u_0$ plane for different values of constant $v_1$ are shown in (b), (d) and (f). In (b) on the left branch a pair of TCPs emerge (shown at $v_2=13.24$, $u_{0c}=0.185$ and $v_2=18.46$, $u_{0c}=0.174$ in the figure). In (d) and (f) the segment of DT line connecting the TCPs lengthen. An island of $r_*>0$ phase is present near the peak $u_0=0.5$ in all the figures (b), (d) and (f). Here we used $D=1$, $x_m=0.5$ and $L=1$.}
   \label{fig:Model3_tran_line}
   \end{minipage}
  \hfill
  \begin{minipage}{0.4\textwidth}
  \end{minipage}
\end{figure}

The above scenario changes as $v_1$ is decreased, and $v_2> -v_1$, such that the potential starts to develop a peak at $x_m$.  While for $v_2<0$ the transitions in Fig.(\ref{fig:Model3_tran_line2}), Fig.(\ref{fig:Model3_tran_line3}) and Fig.(\ref{fig:Model3_tran_line4}) remain similar to Fig.(\ref{fig:Model3_tran_line1}), the ORRVT lines in the $v_2-u_0$ plane start to develop completely new features for $v_2>0$ regime. In Figs.(\ref{fig:Model3_Pot_line2},\ref{fig:Model3_Pot_line3},\ref{fig:Model3_Pot_line4}) we systematically reduce $v_1$ from positive to negative values. In Fig.(\ref{fig:Model3_tran_line2}) for $v_2>0$, we see the appearance of the first new feature --- the critical line on the left side splits by producing a pair of TCPs. As $v_1$ is further lowered we see in Fig.(\ref{fig:Model3_tran_line3}) and Fig.(\ref{fig:Model3_tran_line4}) that the DT line connecting the TCPs stretch out in length. A second new feature in the Figs.(\ref{fig:Model3_tran_line2},\ref{fig:Model3_tran_line3},\ref{fig:Model3_tran_line4}), is the appearance of an island like domain near the peak, where resetting is beneficial ($r_*>0$). Thus at a large enough value of $v_2$ on varying $u_0$, one would encounter four CT points.    

\begin{figure}[ht!]
  \begin{subfigure}{0.225\textwidth}
      \vspace*{0.0cm}
    \hspace*{-0.5cm}
    \vspace*{0.0cm}
    \includegraphics[width = 1.1\textwidth,height=0.12\textheight]{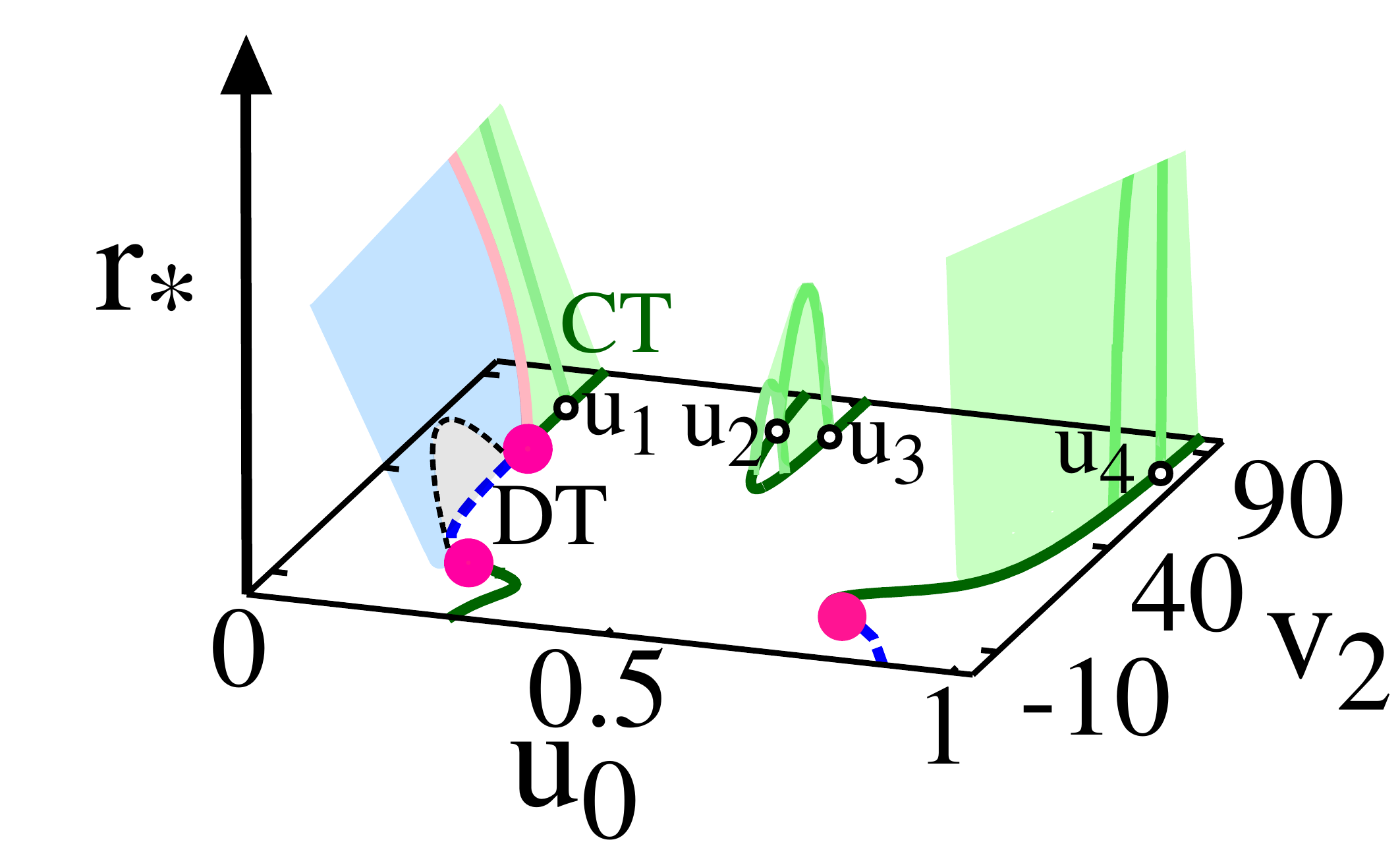}
   \caption[]{} 
    \label{fig:Model3_tran_line11}
  \end{subfigure}
  \begin{subfigure}{0.225\textwidth}
    \includegraphics[width = 1\textwidth,height=0.12\textheight]{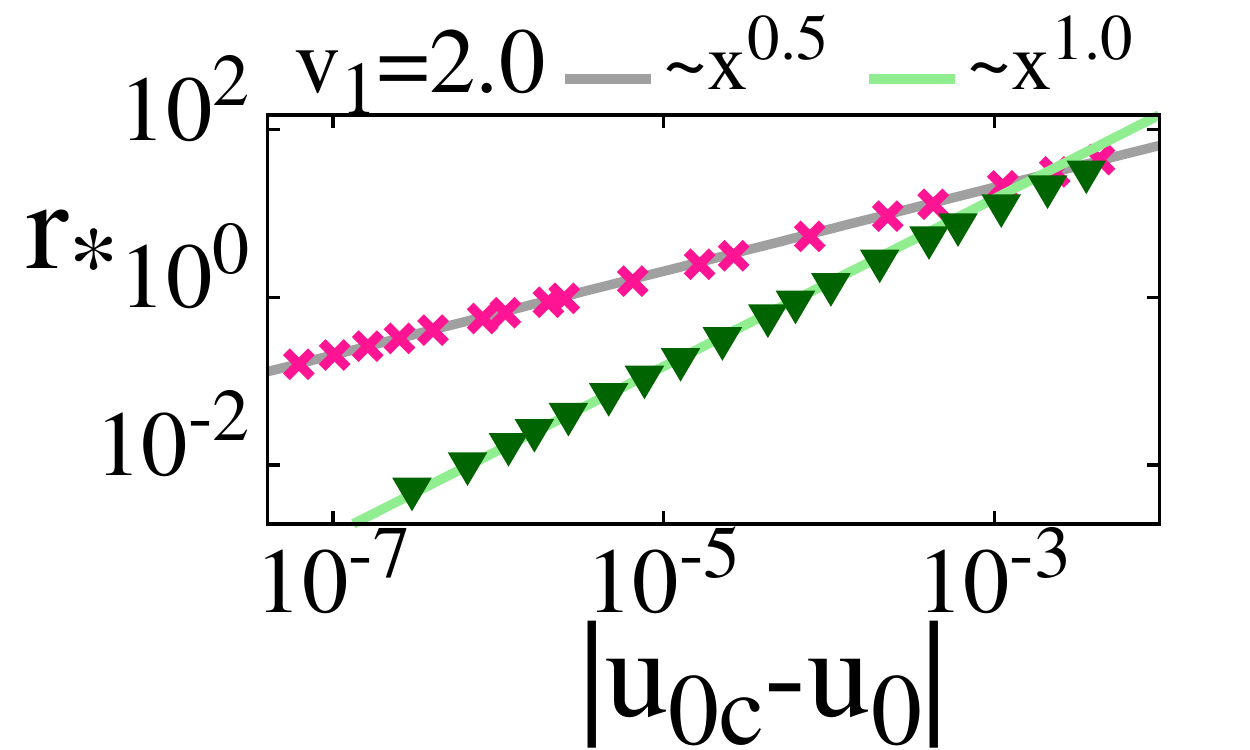}
   \caption[]{} 
    \label{fig:Model3_tran_line22}
   \end{subfigure}
  \caption{(a) We show $r_*$ as a function of $v_2$ and $u_0$, for $v_1=2$. The $r_*$ surface in blue jumps abruptly to $r_*=0$ at the DT line between the two TCPs (shown in pink circles at $u_{\text{TCP}}=0.224$ and $0.149$). In contrast the $r_*$ surface (shown in green) and the line in purple, go to $r_*=0$ continuously at the CT line and TCP point respectively. At $v_2= 75$ we have four CT points: $u_1=0.149$, $u_2=0.449$, $u_3=0.513$ and $u_4=0.973$. The value of $r_*$ rises from $0$ at $u_2$ reaches a peak and comes down and vanishes at $u_3$. The CT line near $u_0=1$ continues to a DT line through a TCP. In (b) we show in log-log plot of the power-laws that $r_*$ follows on approaching $u_{0c}=u_1$ and $u_{0c}=u_{\text{TCP}}=0.149$ with exponents $\beta=1$ and $1/2$ respectively. }
    \label{fig:Model3_ORR_T}
\end{figure}

We show the variation of the ORR value $r_*$ as a function of $v_2$ and $u_0$ for $v_1=2$ in Fig.(\ref{fig:Model3_tran_line11}). At high value of $v_2$ we have marked the four transition points of ORR namely $u_1$, $u_2$, $u_3$ and $u_4$ which would be encountered if one varies $u_0$ at $v_2=75$. The regions of $r_*>0$ (shaded in color) are interspersed with regions of $r_*=0$ (shown in white). The discontinuous jump in $r_*$ is also shown over a stretch of the transition line near the left boundary. This first order stretch is terminated by two TCPs at the two ends, which is a rather rare feature.  In Sec.[\ref{sec2}], we have analytically established that the ORR behaves as $r_*\propto [\lambda_c-\lambda]^{\beta}$ with $\beta=1$ near continuous transition and $\beta=1/2$ near TCP. We see in the Fig.(\ref{fig:Model3_tran_line11}) that the approach of $r_*$ to the critical point ($u_1$) looks linear (green curve), while that to the TCP is a curved line (pink). We explicitly check the exponent $\beta$ in Fig.(\ref{fig:Model3_tran_line22}) in a log-log plot of $r_*$ vs. $|u_{0c}-u_0|$. At the CT point  $u_{0c}=u_1$ we have $\beta=1$ and at the TCP $u_{0c}=u_{\text{TCP}}$ we have $\beta=1/2$, as expected.
The plots in Figs.(\ref{fig:Model3_tran_line}) and (\ref{fig:Model3_ORR_T}) are for finite values of $v_2$. A natural question is whether the scenarios presented for ORRVT lines continue to $v_2\to \pm\infty$ or not. By setting the condition  $\langle T^2 \rangle_{-}=2\langle T \rangle^2_{-}$ for $x<x_m$  at $v_2\to +\infty$ we obtain the Eq.(\ref{M3_v2_infty_left}) in App.[\ref{App_M3}]. This transcendental equation has three real solutions --- for example for $v_1=2$ the two critical points are $u_{1}=0.149$, $u_{2}=0.376$ and $u_3=0.5$. Similarly, for $x>x_m$ setting $\langle T^2 \rangle_{+}=2\langle T \rangle^2_{+}$ as $v_2\to\infty$ we find that the critical point is $u_{4}\to 1$ independent of the value of $v_1$  (see Eq. (\ref{M3_v2_infty_right}) in App.[\ref{App_M3}]). Thus we have proved that four critical points continue to exist at $v_2\to \infty$.

On the other hand, when $v_2\to -\infty$ by setting $\langle T^2 \rangle_{-}=2\langle T \rangle^2_{-}$ for $x<x_m$ we obtain the transcendental Eq.(\ref{M3_v2_n_infty_left}) in  App.[\ref{App_M3}], which has only a single real solution --- for example $u_{0c}=0.260$ for $v_1=2$. In contrast $\langle T^2 \rangle_{+} \gg 2\langle T \rangle^2_{+}$ and hence there
is no solution for a CT. Thus for $x>x_m$ there is no CT, although  we know there is a DT (from Fig.(\ref{fig:Model3_tran_line})). Thus we have proved that as $v_2\to -\infty$, there 
is only a single CT. 

\begin{figure}
   \begin{minipage}{0.5\textwidth}
     \begin{subfigure}{0.45\textwidth}
        \vspace*{0.05cm}
    \hspace*{-0.5cm}
    \vspace*{0.11cm}
   \includegraphics[width = 1.1\textwidth,height=0.11\textheight]{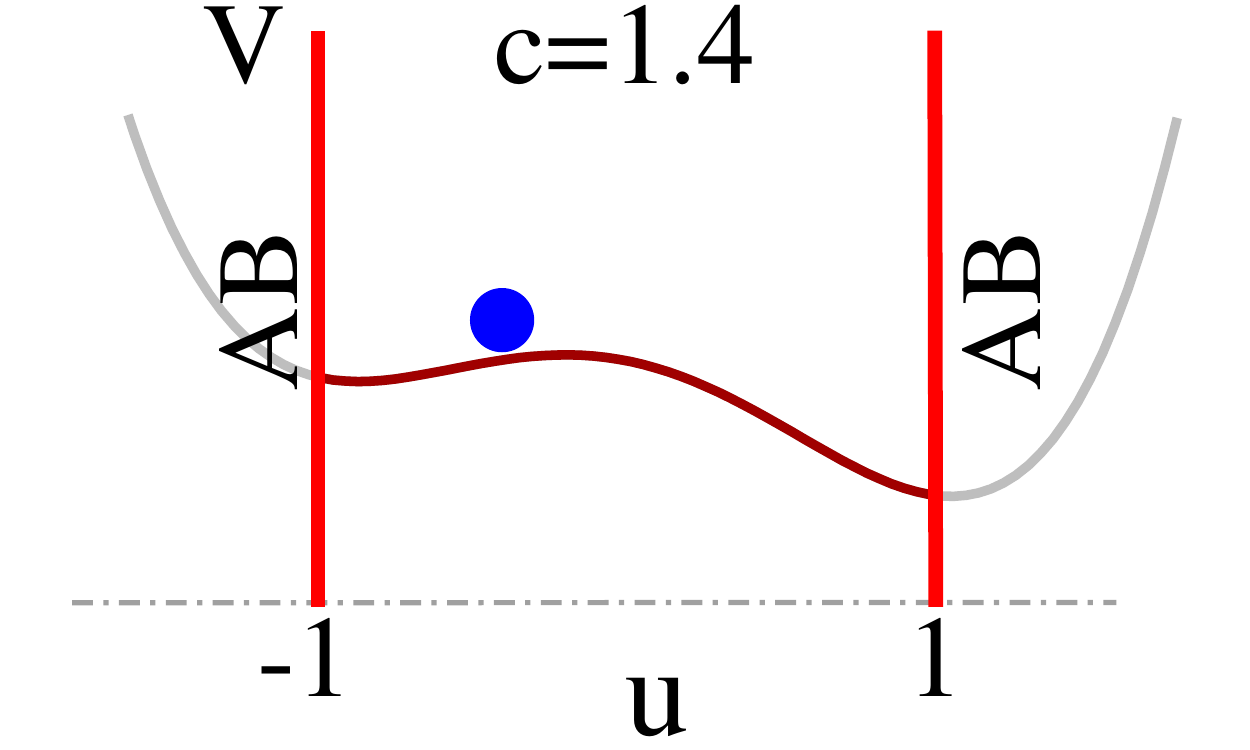}
   \caption[]{} 
    \label{fig:Model4_Pot_line1}
 \end{subfigure}
\begin{subfigure}{0.45\textwidth}
    \includegraphics[width = 1.1\textwidth,height=0.12\textheight]{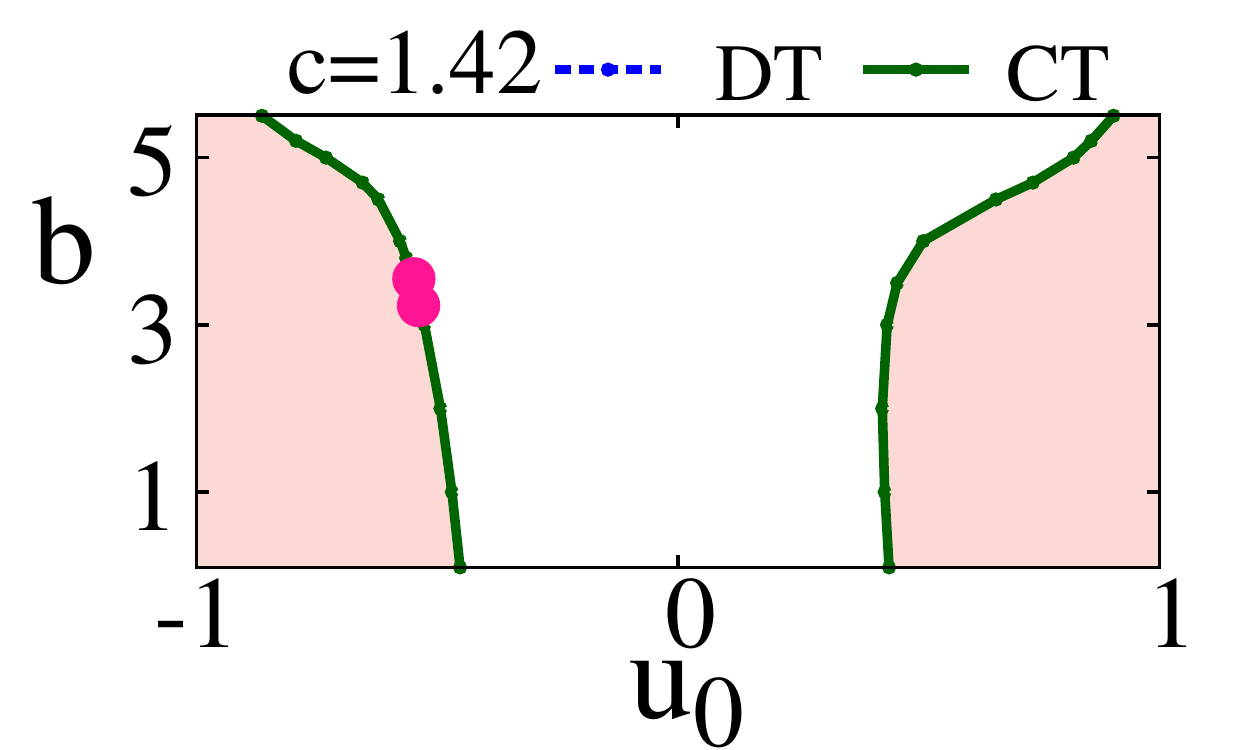}
   \caption[]{} 
    \label{fig:Model4_tran_line1}
   \end{subfigure}
\begin{subfigure}{0.45\textwidth}
   \vspace*{0.1cm}
    \hspace*{-0.5cm}
    \vspace*{0.11cm}
    \includegraphics[width =1.1\textwidth,height=0.11\textheight]{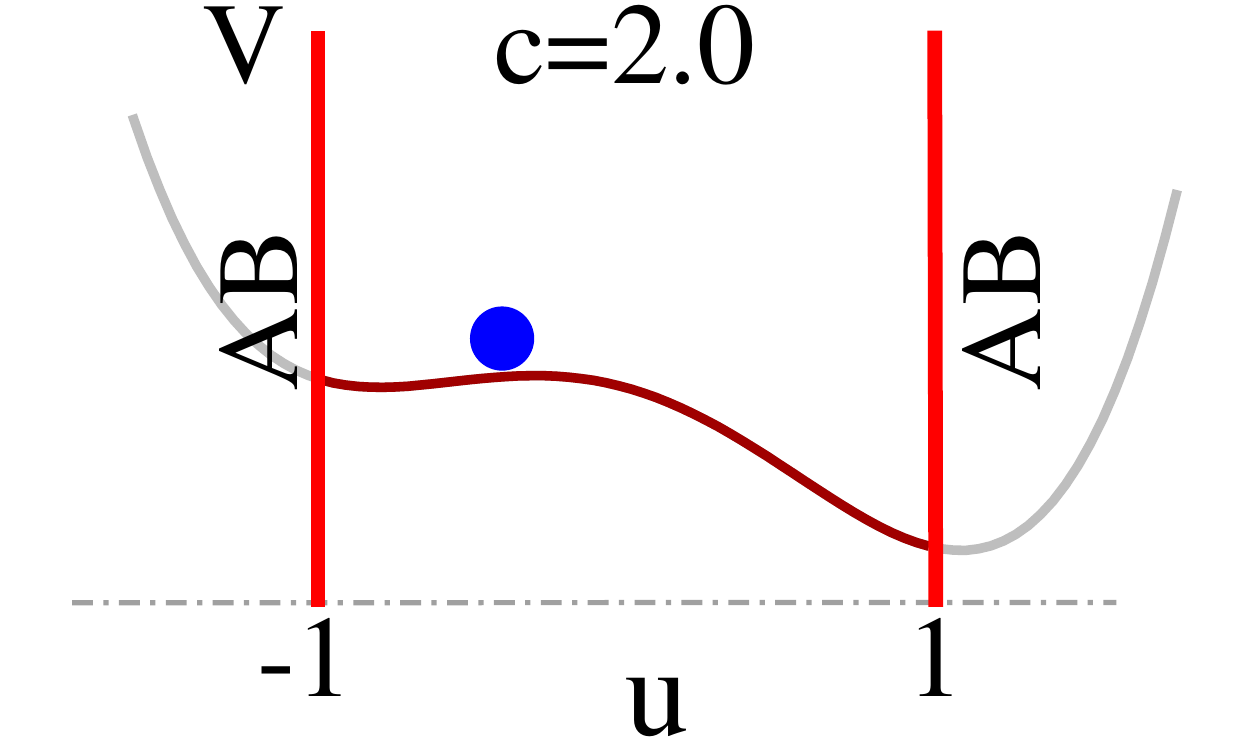}
    \caption[]{} 
    \label{fig:Model4_Pot_line2}
  \end{subfigure}
  \begin{subfigure}{0.45\textwidth}
    \includegraphics[width = 1.1\textwidth,height=0.12\textheight]{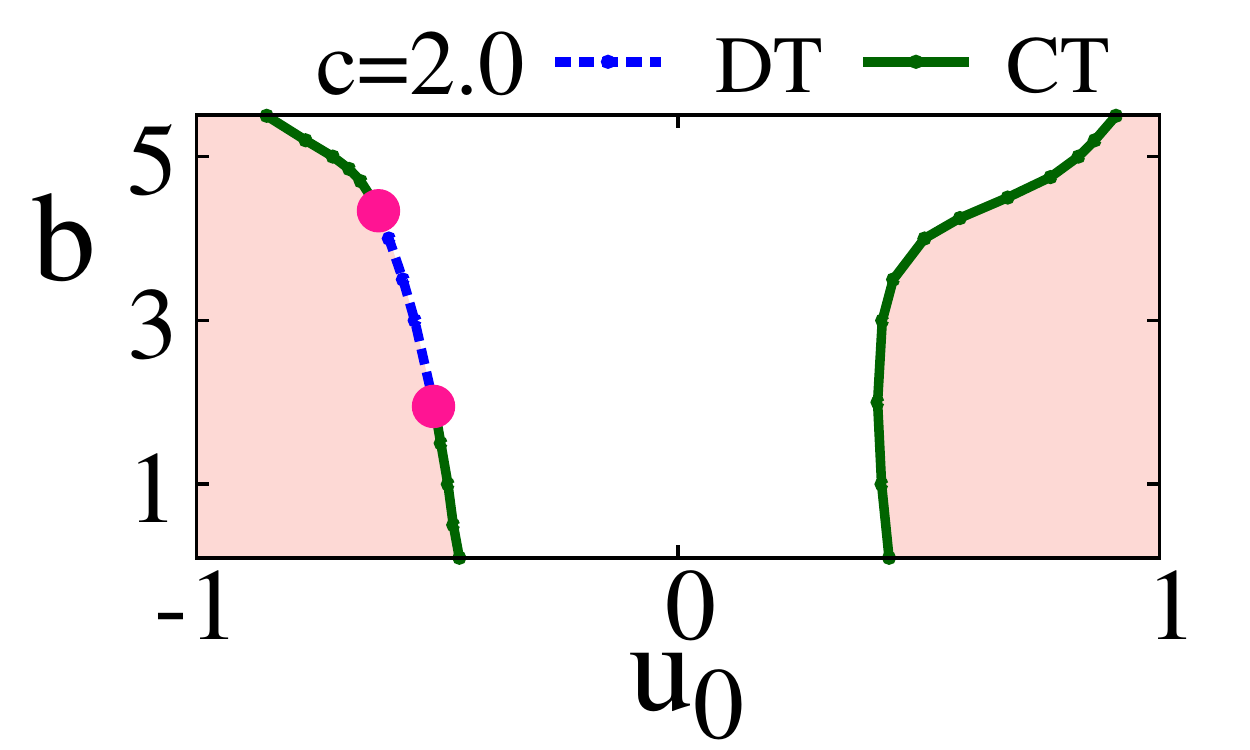}
   \caption[]{} 
    \label{fig:Model4_tran_line2}
   \end{subfigure}
  \begin{subfigure}{0.45\textwidth}
     \vspace*{0.1cm}
    \hspace*{-0.5cm}
    \vspace*{0.11cm}
  \includegraphics[width = 1.1\textwidth,height=0.11\textheight]{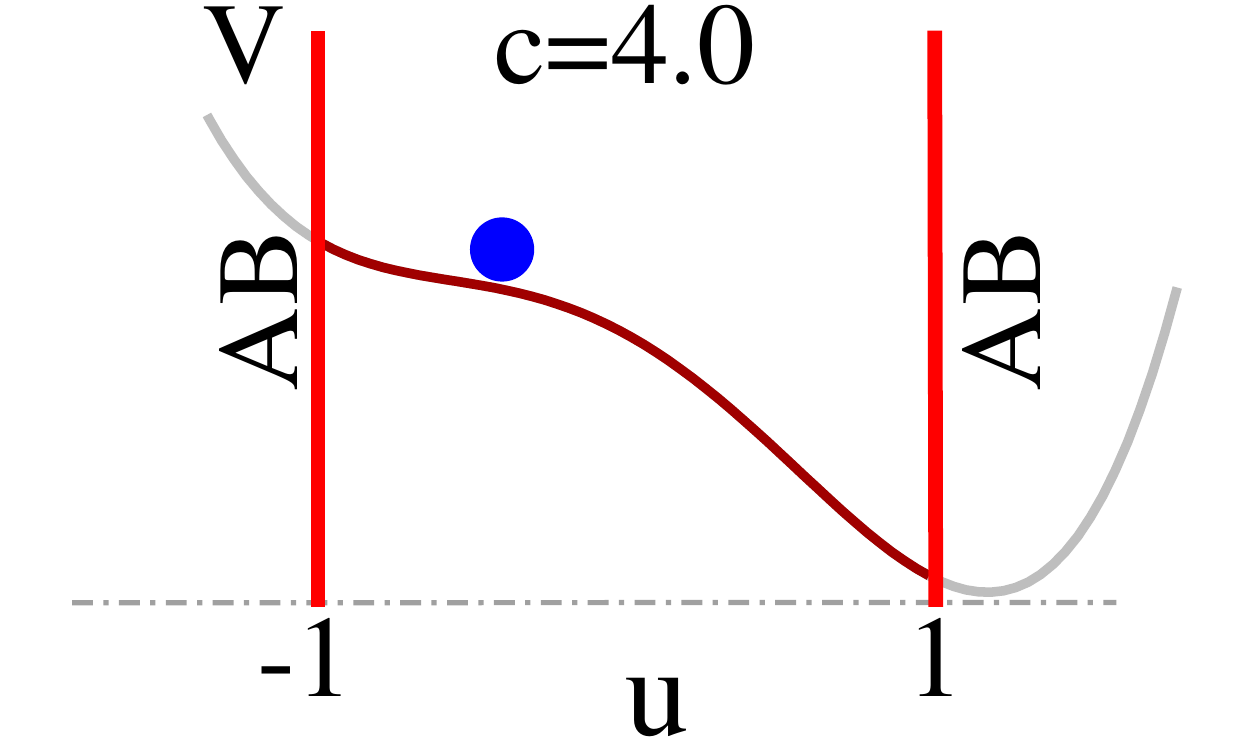}
   \caption[]{} 
    \label{fig:Model4_Pot_line3}
 \end{subfigure}
 \begin{subfigure}{0.45\textwidth}
    \includegraphics[width = 1.1\textwidth,height=0.12\textheight]{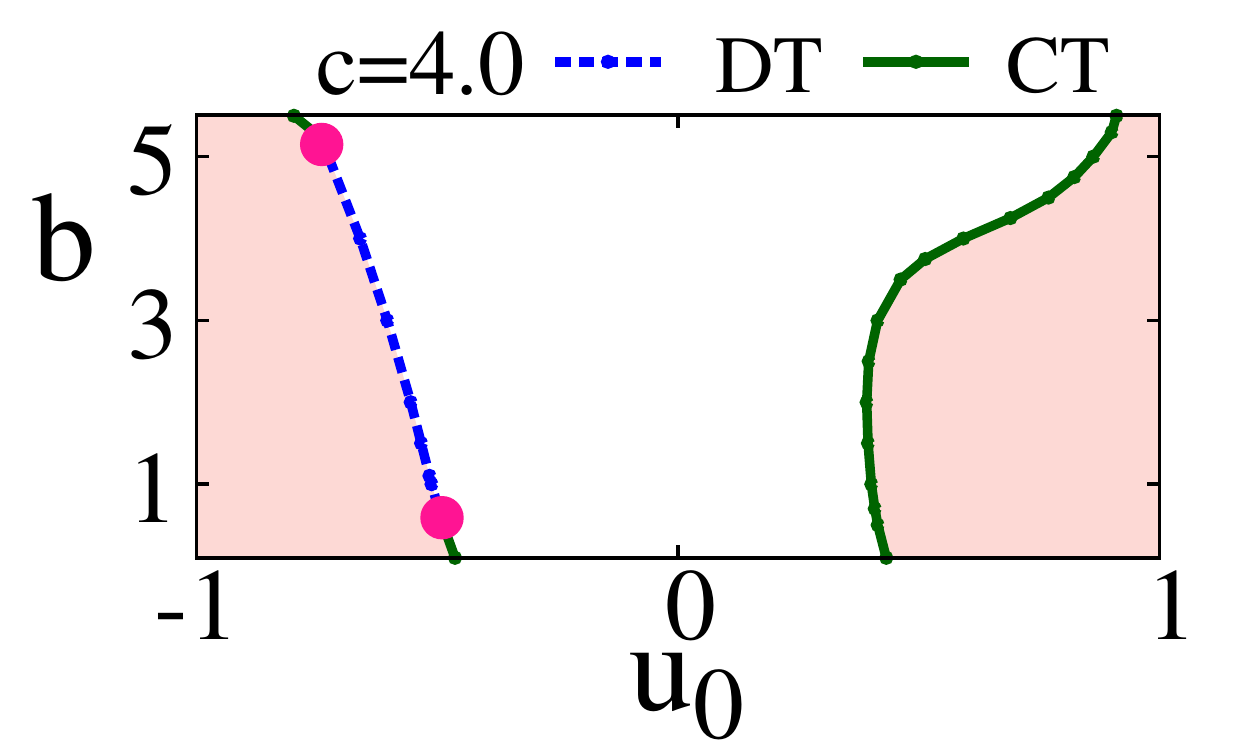}
    \caption[]{} 
    \label{fig:Model4_tran_line3}
   \end{subfigure}
 \caption{The shape of the quartic potential within the allowed domain $u\in[-1,1]$ is shown in dark line for $b=3$ and three different magnetic fields: (a) $c=1.4$, (c)  $c=2$, and (e) $c=4$. The corresponding ORRVT lines in the $b-u_0$ plane are shown in (b), (d), and (f) respectively. Observe the lengthening of the DT segment between two TCPs (pink circles) as the tilt of the potential increases with increasing $c$. Here $D=6T/T_c$.}
 \label{fig:Model4_tran_line}
 \end{minipage}
  \hfill
  \begin{minipage}{0.4\textwidth}
  \end{minipage}
\end{figure}

\subsection{Model-IV}
In Model-III the slopes $v_1$ and $v_2$ can be independently varied from positive to negative values, and thus the depth of any one valley and the sharpness of the peak can be independently tuned. In comparison in Model-IV, changing $b$ affects the depth of both the valleys, while changing $c$ affects the asymmetry of both the valleys simultaneously. Thus fine tuning the shape of the potential due to which the interesting features of the phase diagram in Fig.(\ref{fig:Model3_tran_line3}) arise, becomes rather challenging to be achieved in Model-IV. One may obtain some of the results similar to Model-III, but not all, as shown below.

 We obtain $\langle T_r \rangle$ by following the method discussed in Sec.[\ref{numerical_methos}] with the boundary conditions $\langle T_r \rangle=0$ at $u=\pm1$.  The minimum value of $\langle T_r \rangle$ gives us the ORR value $r_*$. As in Model-II, the magnetization evolves via a temperature dependent diffusion constant $D\propto T/T_c$ in between successive resets. In Figs.(\ref{fig:Model4_Pot_line1},\ref{fig:Model4_Pot_line2},\ref{fig:Model4_Pot_line3}), we show the shape of the potential $V(x)$ as the magnetic field $c$ is varied from small to large values at $b=3$. We see that the critical line in the $b-u_0$ plane near the left AB, splits into a pair of TCPs in Fig.(\ref{fig:Model4_tran_line1}). The critical line near the right AB is one of a CT. In Figs.(\ref{fig:Model4_tran_line2},\ref{fig:Model4_tran_line3}), the distance between the TCPs connected by the (dashed) line of DT lengthen. This feature resembles closely what we saw in Figs.(\ref{fig:Model3_tran_line2},\ref{fig:Model3_tran_line3},\ref{fig:Model3_tran_line4}). But we do not see an island like region with $r_*>0$ near the potential peak. This is understandable as we cannot have high values of $b$ in this system due to physical constraints. One cannot have a temperature below $T=0$, i.e. we cannot make $b>6$. Also changing $c$ to $-c$ would lead to swapping the left and the right ORRVT branches in the $b-u_0$ plane --- thus we do not expect any further new feature to arise, than shown in Fig.(7).

\section{Conclusion}
In this paper, we studied different models of $1-d$ diffusion, to understand whether the strategy of stochastic resetting is beneficial to reach target(s), in the presence of spatially varying potentials. In particular, we focused on potentials having a local peak or barrier with valleys on the two sides, flanked by reflecting or absorbing boundaries. The Models-I and III had piecewise linear potentials and hence were analytically exactly tractable. The Models-II and IV described the time evolution of magnetization of a magnetic system subjected to stochastic resetting --- the realistic quartic potential in this case permitted us to do numerical study.

The subtle interplay among features like the sharpness of the peak, the asymmetric depth of the valleys, and the presence of either RB or an AB, in the different models produced spatially interspersed domains where resetting is either beneficial or harmful. The boundary of these domains were maked by continuous or discontinuous ORRVT. In some of the models we found multiple tri-critical points. 

For Model-I and Model-II with an AB and a RB,  we obtain an interesting result that for intermediate sharpness of the peak, there are two neighborhoods (near the AB and near the peak) where resetting is beneficial and the remaining  two neighborhoods where it is not. Consequently, there are three continuous transition points.

For Model-III with two ABs we find two novel features. The first is the existence of a discontinuous transition line flanked by two TCPs on the two sides. One TCP where the first order and a second order line meets has been found in earlier literature. Compared to that, the existence of two TCPs with a finite discontinuous transition segment is exotic. The second feature is the existence of four critical points and associated three domains in space (two near the ABs and another around the barrier) where resetting is beneficial.

One important understanding that we have from this work is that discontinuous ORRVT can only arise if we have more than one ABs. Additionally, there should be space dependent asymmetric potential in the region of confinement. Recall that, we had discontinuous transitions and TCPs in Models-III and IV and not in Models-I and II. The potential barriers enhance the necessity of resetting in their neighborhood.   

We note that earlier studies on first passage with resetting from a valley to a hill top did not have such varied types and number of ORR vanishing transitions. What we have shown is that having the full spatial variation of the potential with both valleys and a barrier, richer aspects of ORRVT may arise.  

The fact that we considered thermal evolution of the magnetic system in Model-II and IV with the appropriate temperature dependent diffusion constant, in between two resets, make the results obtained relevant to realistic systems. In particular it would be nice to do simulation of a many body Ising magnetic system with stochastic reset to confirm whether the results we obtained for an effective one body problem remain valid.

{\bf Acknowledgement}: DD would like to acknowledge SERB India (grant no. MTR/2019/000341) for financial support.

\appendix
\begin{widetext}
 \label{Appendix}
\section{\label{App_sec1}Matching conditions for the tent potential at $x=x_m$}
Here we discuss, the matching conditions for the peicewise solutions of the survival probabilities $Q_-(x,t)$ and $Q_+(x,t)$ over the two intervals $[0,x_m)$ and $(x_m,L]$ across $x=x_m$. These are used in calculations of Models-I and III.

The two piecewise backward equations (\ref{eq_BFPE_1}) and (\ref{eq_BFPE_2}) may be combined into a single equation as follows: 
\begin{equation}
\frac{\partial Q(x,t)}{\partial t}=D\frac{\partial^2 Q(x,t)}{\partial x^2}-(v_1-(v_2+v_1)\Theta(x-x_m)) \frac{\partial Q(x,t)}{\partial x}-rQ(x,t)+rQ(x_{0},t)
\label{eq_BFPE_3}
\end{equation}
where, $\Theta$ represents Heaviside unit step function \cite{Arfken_7ed_book}.

Integrating both sides of Eq. (\ref{eq_BFPE_3}) from $x=x_m-\epsilon$ to $x=x_m+\epsilon$, where $\epsilon$ is an infinitesimally small positive number, we have:
\begin{align}
\nonumber\int_{x_m-\epsilon}^{x_m+\epsilon}dx \frac{\partial Q(x,t)}{\partial t}=&\int_{x_m-\epsilon}^{x_m+\epsilon}dxD\frac{\partial^2 Q(x,t)}{\partial x^2}-\int_{x_m-\epsilon}^{x_m+\epsilon}dx(v_1-(v_2+v_1)\Theta(x-x_m)) \frac{\partial Q(x,t)}{\partial x}\\
&-\int_{x_m-\epsilon}^{x_m+\epsilon}dxrQ(x,t)+\int_{x_m-\epsilon}^{x_m+\epsilon}dxrQ(x_{0},t)
\label{eq_BFPE_4}
\end{align}

Since $Q(x,t)$ and $\frac{\partial Q(x,t)}{\partial t}$ are finite everywhere, it implies $\int_{x_m-\epsilon}^{x_m+\epsilon}Q(x,t)dx=2\epsilon Q(x_m,t)\rightarrow 0$, $\int_{x_m-\epsilon}^{x_m+\epsilon}dx \frac{\partial Q(x,t)}{\partial t}=2\epsilon  \frac{\partial Q(x_m,t)}{\partial t}\to 0$ and $\int_{x_m-\epsilon}^{x_m+\epsilon}dxQ(x_{0},t)\to 0$. Then Eq. (\ref{eq_BFPE_4}) reduces to 
\begin{align}
\int_{x_m-\epsilon}^{x_m+\epsilon}dxD\frac{\partial^2 Q(x,t)}{\partial x^2}
  &=\int_{x_m-\epsilon}^{x_m+\epsilon}dx(v_1-(v_2+v_1)\Theta(x-x_m)) \frac{\partial Q(x,t)}{\partial x}\nonumber\\
&=v_1\int_{x_m-\epsilon}^{x_m}dx\frac{\partial Q(x,t)}{\partial x}-v_2\int_{x_m}^{x_m+\epsilon}dx\frac{\partial Q(x,t)}{\partial x}
\label{eq_BFPE_5}
\end{align}
This gives
\begin{align}
  D\frac{\partial Q(x,t)}{\partial x}\Bigr|_{x=x_m+\epsilon}-\frac{\partial Q(x,t)}{\partial x}\Bigr|_{x=x_m-\epsilon}&=v_1(Q(x_m,t)-Q(x_m-\epsilon,t))\nonumber\\
  &-v_2(Q(x_m+\epsilon,t)-Q(x_m,t))
\end{align}
Assuming $Q(x,t)$ is continuous through $x=x_m$, the right hand side of the above equation vanishes and we see that the first derivative  $\frac{\partial Q(x,t)}{\partial x}$ is also continuous. Thus we have
\begin{equation}
Q_-(x,t)=Q_+(x,t)
\end{equation}
\begin{equation}
Q^{'}_-(x,t)=Q^{'}_+(x,t)
\end{equation} 
In the Laplace space these matching conditions give:
\begin{align}
  q_-(x,s)=q_+(x,s)\label{App_eq:Lap_survive_Match1}\\  
  q^{'}_-(x,s)=q{'}_+(x,s)
    \label{App_eq:Lap_survive_Match2}
\end{align}
Since the $n^{th}$ moment $\langle T_r^{n} \rangle =n(-1)^{n-1}\frac{\partial^{n-1} q(x_0,s)}{\partial s^{n-1}}\bigg|_{s\to0}$ the matching conditions for the moments are:
\begin{align}
    \langle T_r^{n} \rangle_{-}=\langle T_r^{n} \rangle_{+}\\
    \langle T_r^{n} \rangle^{'}_{-}=\langle T_r^{n} \rangle^{'}_{+}
    \label{App_eq:MFPT_Match}
\end{align}

\section{\label{App_sec2_tent-pot}Exact Moments for the Tent-Potential to study CT and DT in Model-I and Model-III}
In this part we find the first two moments without resetting ($r=0$) to study the CT using the criterion Eq.(\ref{Conti_Trans}). Then we obtain the MFPT $\langle T_r\rangle$ with resetting ($r \neq 0$) to study the DT.
\subsection{\label{App_sec2_M1}Model-I}
\par\texttt{Moments without resetting}: For $x<x_m$ we solve Eq.(\ref{diff_tent_Mom_l}) and for $x>x_m$ we solve Eq.(\ref{diff_tent_Mom_R}) with the matching and boundary conditions discussed in Sec[\ref{Model1_discus}] to obtain $\langle T\rangle_{-}, \langle T^2\rangle_{-}, \langle T\rangle_{+}$ and $\langle T^2\rangle_{+}$ as follows:
\be
\langle T\rangle_{-}=\frac{x}{v_1}+\frac{D \left(e^{\frac{v_1 x}{D}}-1\right) e^{-\frac{\left(v_1+v_2\right) x_m}{D}} \left(v_1 e^{\frac{v_2L}{D}}-\left(v_1+v_2\right) e^{\frac{v_2 x_m}{D}}\right)}{v_1^2 v_2}
\label{eq:Model1_1st_m_left}
\ee
\begin{equation} 
\begin{split}
 \langle T^2 \rangle_{-}
&=\frac{1}{v_1^4 v_2^3} \bigg[ 2 D^2 \left(e^{\frac{v_1 x}{D}}-1\right) e^{-\frac{3 \left(v_1+v_2\right) x_m}{D}} \bigg(v_1^2 \left(\left(v_1+v_2\right) e^{\frac{v_1 x_m}{D}}-v_2\right) e^{\frac{2 L v_2+\left(v_1+v_2\right) x_m}{D}}\\
  &\quad-2 v_2 v_1 \left(v_1 \left(e^{\frac{v_1 x_m}{D}}-1\right)-v_2\right) e^{\frac{v_2 \left(L+2 x_m\right)+v_1 x_m}{D}}-\left(v_1+v_2\right) e^{\frac{\left(v_1+3 v_2\right) x_m}{D}} \bigg((v_1^2\\
  &\quad-2 v_2 v_1+2 v_2^2) e^{\frac{v_1 x_m}{D}}+v_2 \left(v_1+v_2\right)\bigg)\bigg) +2 D v_1 v_2 e^{-\frac{\left(v_1+v_2\right) x_m}{D}} \bigg(-2 v_1^2 e^{\frac{L v_2}{D}}\\
  &\quad \left(e^{\frac{v_1 x}{D}}-1\right) \left(L-x_m\right)-v_2 v_1 \big(\left(x-2 x_m\right) e^{\frac{v_1 x}{D}}+2 x_m+x\big) \left(e^{\frac{L v_2}{D}}-e^{\frac{v_2 x_m}{D}}\right)\\
&\quad+v_2^2 e^{\frac{v_2 x_m}{D}} \left(x \left(e^{\frac{v_1 x_m}{D}}+e^{\frac{v_1 x}{D}}+1\right)-2 x_m \left(e^{\frac{v_1 x}{D}}-1\right)\right)\bigg)+v_1^2 v_2^3 x^2\bigg]\\
\end{split}
\label{Mode-I_2nd_m_left}
\end{equation}
\begin{equation}
\begin{split}
 \langle T\rangle_{+}
  &=\frac{D\left( \left(v_1+v_2\right) \bigg(v_1 e^{\frac{v_2(L - x_m)}{D}}+v_2e^{-\frac{v_1 x_m}{D}}\bigg)-v_1v_2e^{\frac{v_2(L-x_m)-v_1x_m}{D}}\right)}{v_1^2 v_2^2}\\
 &\quad+\frac{v_2 \left(v_1+v_2\right)\left(v_1 x_m-D\right)}{v_1^2 v_2^2}-\frac{D e^{\frac{v_2 (L-x)}{D}}}{v^2_2}-\frac{x}{v_2}\\
\end{split}
\label{eq:Model1_1st_m_right}
\end{equation}
\begin{equation} 
\begin{split}
\langle T^2 \rangle_{+}
&=\frac{e^{-\frac{v_1 x_m+v_2 \left(x+x_m\right)}{D}} }{v_1^2 v_2^4} \bigg[ e^{\frac{v_1 x_m+v_2 \left(x+x_m\right)}{D}} v_2^2 \left(x v_1-\left(v_1+v_2\right) x_m\right)^2+\frac{1}{v_1^2}\bigg[2 D^2 e^{-\frac{\left(v_1+v_2\right) x_m}{D}}\\
  &\quad\bigg(e^{\frac{L v_2+2 v_1 x_m}{D}} \big(e^{\frac{x v_2}{D}}-e^{\frac{v_2 x_m}{D}}\big) \left(e^{\frac{L v_2}{D}}+2 e^{\frac{v_2 x_m}{D}}\right) v_1^4-e^{\frac{v_1 x_m}{D}} \left(1-e^{\frac{v_1 x_m}{D}}\right) \left(e^{\frac{L v_2}{D}}-e^{\frac{v_2 x_m}{D}}\right)\\
  &\quad\bigg(2 e^{\frac{(L+x) v_2}{D}}-e^{\frac{v_2 \left(L+x_m\right)}{D}}+e^{\frac{v_2 \left(x+x_m\right)}{D}}\bigg) v_2 v_1^3-\left(1-e^{\frac{v_1 x_m}{D}}\right)\bigg(e^{\frac{v_1 x_m}{D}} \bigg(e^{\frac{(2 L+x) v_2}{D}}+e^{\frac{v_2 \left(L+2 x_m\right)}{D}}\\
  &\quad-3 e^{\frac{v_2 \left(L+x+x_m\right)}{D}}+e^{\frac{v_2 \left(x+2 x_m\right)}{D}}\bigg)-e^{\frac{(2 L+x) v_2}{D}}+2 e^{\frac{v_2 \left(L+x+x_m\right)}{D}}-e^{\frac{v_2 \left(x+2 x_m\right)}{D}}\bigg)v_2^2 v_1^2-2\\
  &\quad\left(1-e^{\frac{v_1 x_m}{D}}\right)\left(e^{\frac{v_2 \left(L+x+x_m\right)}{D}}-e^{\frac{v_2 \left(x+2 x_m\right)}{D}}\right) v_2^3 v_1+e^{\frac{v_2 \left(x+2 x_m\right)}{D}} \left(1+e^{\frac{v_1 x_m}{D}}-2 e^{\frac{2 v_1 x_m}{D}}\right) v_2^4\bigg)\bigg]\\
  &-\frac{1}{v_1}\bigg[2 D e^{-\frac{2 \left(v_1+v_2\right) x_m}{D}} v_2 \bigg(e^{\frac{L v_2+3 \left(v_1+v_2\right) x_m}{D}}\big(v_2 x_m+v_1 \left(-2 L+x+x_m\right)\big) v_1^2+e^{\frac{(L+x) v_2+2 \left(v_1+v_2\right) x_m}{D}}\\
  &\quad v_2 \left(3 \left(v_1+v_2\right) x_m-(2 L+x) v_1\right) v_1+e^{\frac{3 v_1 x_m+v_2 \left(L+x+2 x_m\right)}{D}}  \left(v_1 \left(2 L+x-3 x_m\right)-v_2 x_m\right) v_1\\
    &\quad\left(v_1+v_2\right)+e^{\frac{x v_2+3 \left(v_1+v_2\right) x_m}{D}} \bigg(\left(x-x_m\right) v_1^2+v_2 \left(x_m-x\right) v_1-v_2^2 \left(x-2 x_m\right)\bigg) v_1\\
    &\quad-e^{\frac{2 v_1 x_m+v_2 \left(x+3 x_m\right)}{D}} v_2 \left(v_1+v_2\right) \left(3 v_2 x_m+v_1 \left(x_m-x\right)\right)\bigg)\bigg]\bigg]\\
\end{split}
\label{Mode-I_2nd_m_left}
\end{equation}

\par\texttt{MFPT with resetting}:
We solve Eqn.(\ref{eq_lap_BFPE_1}, \ref{eq_lap_BFPE_2}) to find $\langle T_r \rangle_{-}=q_-(x,0)$ and $\langle T_r \rangle_{+}=q_+(x,0)$ as functions of $r$ by using matching conditions Eqn.(\ref{App_eq:Lap_survive_Match1}), (\ref{App_eq:Lap_survive_Match2}) at $x=x_m$ and boundary conditions $q_-(x,s)|_{x=0}=0$, $q'_+(x,s)|_{x=L}=0$. The results are as follows:
\be
\begin{split}
\langle T_r \rangle_{-}
&= \bigg[\alpha ^+ \beta ^+ \left(1-e^{\alpha ^- x_0}\right) e^{\beta ^+ L+\left(\alpha ^++\beta ^-\right) x_m}+\beta ^- \bigg(\beta ^+ \left(e^{\beta ^- L+\beta ^+ x_m}-e^{\beta ^+ L+\beta ^- x_m}\right)\nonumber\\
  &\left(-e^{\alpha ^- x_m}+e^{\alpha ^+ x_m}+e^{\alpha ^- x_m+\alpha ^+ x_0}-e^{\alpha ^+ x_m+\alpha ^- x_0}\right)-\alpha ^+ \left(1-e^{\alpha ^- x_0}\right) e^{\beta ^- L+\left(\alpha ^++\beta ^+\right) x_m}\bigg)\nonumber\\
  & +\alpha ^- \left(1-e^{\alpha ^+ x_0}\right) \left(\beta ^- e^{\beta ^- L+\left(\alpha ^-+\beta ^+\right) x_m}-\beta ^+ e^{\beta ^+ L+\left(\alpha ^-+\beta ^-\right) x_m}\right)\bigg]\bigg/\bigg[r\bigg(-\beta ^- \left(\alpha ^+-\beta ^+\right)\nonumber\\
  & e^{\beta ^- L+\left(\alpha ^++\beta ^+\right) x_m+\alpha ^- x_0}+\beta ^+ \bigg(e^{\beta ^+ L+\beta ^- x_m} \left(\beta ^- e^{\alpha ^- x_m+\alpha ^+ x_0}+\left(\alpha ^+-\beta ^-\right) e^{\alpha ^+ x_m+\alpha ^- x_0}\right)\nonumber\\
  &-\beta ^- e^{\beta ^- L+\left(\alpha ^-+\beta ^+\right) x_m+\alpha ^+ x_0}\bigg)+\alpha ^- e^{\alpha ^+ x_0} \bigg(\beta ^- e^{\beta ^- L+\left(\alpha ^-+\beta ^+\right) x_m}\nonumber\\
 & -\beta ^+ e^{\beta ^+ L+\left(\alpha ^-+\beta ^-\right) x_m}\bigg)\bigg)\bigg]\nonumber\\
\end{split}
\ee
\be
\begin{split}
\langle T_r \rangle_{+}
&= \bigg[\alpha ^- e^{-\alpha ^+ x_m} \left(\beta ^- e^{\beta ^- L} \left(e^{\beta ^+x_m}-e^{\alpha ^+ x_m+\beta ^+ x_0}\right)-\beta ^+ e^{\beta ^+ L} \left(e^{\beta ^- x_m}-e^{\alpha ^+ x_m+\beta ^-x_0}\right)\right) \nonumber\\
& +e^{-\left(\alpha ^-+\alpha ^+\right) x_m} \bigg(\alpha ^+ \beta ^+ e^{\beta ^+ L+\alpha ^+x_m} \left(e^{\beta ^- x_m}-e^{\alpha ^- x_m+\beta ^- x_0}\right)+\beta ^- \bigg(\beta ^+ \left(e^{\alpha ^- x_m}-e^{\alpha ^+x_m}\right)\nonumber\\
  &\left(e^{\beta ^+ L+\beta ^-x_m}-e^{\beta ^- L+\beta ^+x_m}\right)+\alpha ^+ e^{\beta ^- L} \left(e^{\left(\alpha ^-+\alpha ^+\right) x_m+\beta ^+ x_0}-e^{\left(\alpha ^++\beta ^+\right)x_m}\right)\bigg)\bigg)\bigg]\bigg/\bigg[r\left(\alpha ^--\alpha ^+\right)\nonumber\\
  & \left(\beta ^- e^{\beta ^- L+\beta ^+ x_0}-\beta ^+ e^{\beta ^+ L+\beta ^- x_0}\right)\bigg]\nonumber\\
\end{split}
\ee

\par\texttt{Absence of DT in Model-I}: For finite values of $v_1$, $v_2$ and $u_0$ by studying $\langle T_r \rangle$ we could not find any DT. Here we also study the parameter $a_2=\frac{1}{6} \langle T^3 \rangle +\langle T \rangle^3-\langle T \rangle\langle T^2 \rangle$ appearing in Eq.(\ref{eq:small_r_exp}). For $v_1=40$ in Fig .(\ref{fig:Model1_a2_vs_u0}), we plot the exact $a_2$ against $u_0$ for different values of $v_2\in(-\infty, +\infty)$. We see that $a_2$ is always positive, and thus DT cannot arise in Model-I.  
\begin{figure}[ht!]
  \centering
  \includegraphics[width = 0.7\textwidth,height=0.3\textheight]{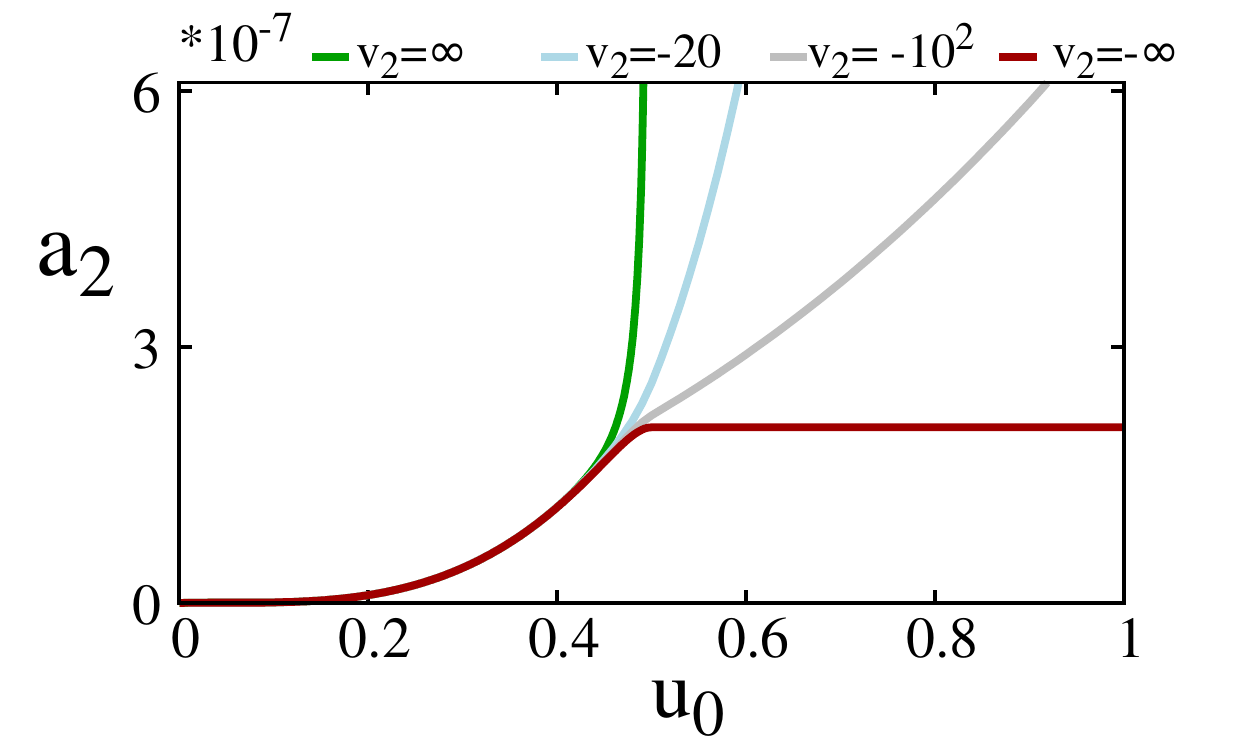}
  \caption{The figure shows variation of $a_2$ with $u_0$ for different values of $v_2$ and at $v_1=40$. We used $D=1$, $x_m=0.5$ and $L=1$.}
    \label{fig:Model1_a2_vs_u0}
\end{figure}

\subsection{\label{App_M3}Model-III}
\par\texttt{Moments without resetting}: For $x<x_m$ and $x>x_m$ we solve Eqn.(\ref{diff_tent_Mom_l}) and (\ref{diff_tent_Mom_R}) respectively with the matching and boundary conditions discussed in Sec[\ref{Model3_discus}] to obtain $\langle T\rangle_{-}, \langle T^2\rangle_{-}, \langle T\rangle_{+}$ and $\langle T^2\rangle_{+}$ as follows:
\be
\begin{split}
\langle T\rangle_{-}
&=\bigg[ D \left(v_1+v_2\right) \left(e^{\frac{v_1 x}{D}}-1\right) \left(e^{\frac{L v_2}{D}}-e^{\frac{v_2 x_m}{D}}\right)+v_2 \bigg(-\left(v_1+v_2\right) x e^{\frac{L v_2+v_1 x_m}{D}}\\
  &\quad+e^{\frac{L v_2+v_1 x}{D}}(\left(v_1+v_2\right) x_m-L v_1)+e^{\frac{L v_2}{D}} \left(L v_1-\left(v_1+v_2\right) x_m+v_2 x\right)+v_1 \\
  &\quad x e^{\frac{\left(v_1+v_2\right) x_m}{D}}\bigg)\bigg]\bigg/\bigg[v_1 v_2\bigg(e^{\frac{L v_2}{D}} \left(v_2-\left(v_1+v_2\right) e^{\frac{v_1 x_m}{D}}\right)+v_1 e^{\frac{\left(v_1+v_2\right) x_m}{D}}\bigg)\bigg]
\end{split}
\label{Model3_First_m_left}
\ee

\be
\begin{split}
\langle T^2 \rangle_{-}
&=\bigg[\frac{e^{-\frac{v_2 x_m}{D}}}{v_1^3 v_2^3 \left(e^{\frac{L v_2}{D}} \left(v_2-\left(v_1+v_2\right) e^{\frac{v_1 x_m}{D}}\right)+v_1 e^{\frac{\left(v_1+v_2\right) x_m}{D}}\right){}^2}\bigg]\bigg[-8 \sinh \left(\frac{x v_1}{2 D}\right) \\
&\quad \sinh \left(\frac{v_2 \left(L-x_m\right)}{2 D}\right)\left(v_1+v_2\right) e^{\frac{v_1 \left(x+x_m\right)+2 v_2 \left(L+2 x_m\right)}{2 D}}\bigg(\sinh \left(\frac{v_2 \left(L-x_m\right)+v_1 x_m}{2 D}\right)\\
  &\quad\left(v_1+v_2\right) \left(v_1^2+v_2 v_1+v_2^2\right)-\sinh \left(\frac{\left(v_1+v_2\right) x_m-L v_2}{2 D}\right)\left(v_1^2-3 v_2 v_1+v_2^2\right)\\
 &\quad\left(v_1-v_2\right) +2 \sinh \left(\frac{v_2 \left(L-x_m\right)}{2 D}\right) \sinh \left(\frac{v_1 x_m}{2 D}\right) \left(v_1^3+v_2^3\right)\bigg) D^2+2 v_2 \bigg(e^{\frac{\left(2 v_1+3 v_2\right) x_m}{D}} \\
  &\quad x v_2^2 v_1^2+e^{\frac{\left(v_1+3 v_2\right) x_m}{D}} v_2 \left(v_1+v_2\right) \left(x+2 x_m\right) v_1^2+e^{\frac{3 v_2 x_m+v_1 \left(x+x_m\right)}{D}} \left(v_1+v_2\right) \left(x-2 x_m\right)\\
  &\quad v_1^2 v_2-2 e^{\frac{L v_2+2 \left(v_1+v_2\right) x_m}{D}} x v_2^2 \left(v_1+v_2\right) v_1-e^{\frac{v_2 \left(L+2 x_m\right)}{D}}\left(3 v_1 \left(x_m-L\right)+v_2 \left(x_m-x\right)\right) \\
  &\quad v_1 v_2 \left(v_1+v_2\right)+e^{\frac{x v_1+v_2 \left(L+2 x_m\right)}{D}} v_1v_2 \left(v_1+v_2\right) (3 v_1 \left(x_m-L\right)+v_2\left(x+x_m\right))-v_1\\
  &\quad e^{\frac{v_1 x_m+v_2 \left(L+2 x_m\right)}{D}} \bigg(v_2 \left(3 L+2 x-x_m\right) v_1^2+v_2^2 \left(L+3 x+4 x_m\right) v_1-v_2^3 \left(x-4 x_m\right)\\
  &\quad +\left(L-x_m\right)v_1^3\bigg) v_1+e^{\frac{v_1 \left(x+x_m\right)+v_2 \left(L+2 x_m\right)}{D}} \bigg(L v_1 \left(v_1^2+3 v_2 v_1+v_2^2\right)-\left(v_1+v_2\right) \big(x_m v_1^2\\
  & \quad+2 x v_2 v_1+v_2^2 \left(x-4 x_m\right)\big)\bigg) v_1+e^{\frac{2 v_1 x_m+v_2 \left(2 L+x_m\right)}{D}} x v_2^2 \left(v_1+v_2\right){}^2+e^{\frac{v_2 \left(2 L+x_m\right)+v_1 \left(x+x_m\right)}{D}}\\
  & \quad\left(v_1+v_2\right) \left(\left(x_m-L\right) v_1^3+v_2 \left(L+x-x_m\right) v_1^2+v_2^2 \left(x-3 x_m\right) v_1-v_2^3 x_m\right)-e^{\frac{v_2 \left(2 L+x_m\right)}{D}}\\
  &\quad v_2^2\left(\left(L+x-2 x_m\right) v_1^2+v_2 \left(x-x_m\right) v_1+v_2^2 \left(x_m-x\right)\right)+e^{\frac{2 L v_2+\left(v_1+v_2\right) x_m}{D}} \left(v_1+v_2\right) \bigg(v_1^3\\
  &\quad \left(L-x_m\right)+v_2 \left(-L+x+x_m\right) v_1^2+v_2^2 \left(x+3 x_m\right) v_1+v_2^3 \left(x_m-2 x\right)\bigg)+e^{\frac{x v_1+v_2 \left(2 L+x_m\right)}{D}}\\
  &\quad v_2^2 \bigg(L v_1^2-\left(v_1+v_2\right) \left(v_1 \left(x+2 x_m\right)-v_2 x_m\right)\bigg)\bigg) D+v_1 v_2^2 \bigg(e^{\frac{2 v_1 x_m+v_2 \left(2 L+x_m\right)}{D}} v_2 \left(v_1+v_2\right){}^2 \\
  &\quad x^2+v_1^2 v_2 x^2 e^{\frac{\left(2 v_1+3 v_2\right) x_m}{D}}-2 e^{\frac{L v_2+2 \left(v_1+v_2\right) x_m}{D}} v_1 v_2 \left(v_1+v_2\right) x^2+e^{\frac{v_2 \left(2 L+x_m\right)}{D}} v_2 \big(L v_1+x v_2\\
  &\quad-\left(v_1+v_2\right) x_m\big){}^2+e^{\frac{v_1 \left(x+x_m\right)+v_2 \left(L+2 x_m\right)}{D}} v_1 \left(\left(v_1+v_2\right) x_m-L v_1\right) \big(-3 L v_1-2 x v_2+3 x_m\\
  &\quad\left(v_1+v_2\right) \big)+ \left(v_1+v_2\right) e^{\frac{v_2 \left(2 L+x_m\right)+v_1 \left(x+x_m\right)}{D}}\left(\left(v_1+v_2\right) x_m-L v_1\right) \big(v_1 \left(x_m-L\right)+v_2 \\
  &\quad\left(2 x-3 x_m\right)\big)-e^{\frac{x v_1+v_2 \left(2 L+x_m\right)}{D}} v_2\left(\left(v_1+v_2\right) x_m-L v_1\right) \left(v_1 \left(x_m-L\right)+v_2 \left(2 x+x_m\right)\right)\\
  &\quad+e^{\frac{2 L v_2+\left(v_1+v_2\right) x_m}{D}} \left(v_1+v_2\right) \bigg(\big(-2 x^2+2 x_m x+3 x_m^2\big) v_2^2-2 v_1 \left(L-x_m\right) \left(x+x_m\right) v_2\\
  &\quad -v_1^2 \left(L-x_m\right){}^2\bigg)-e^{\frac{v_1 x_m+v_2 \left(L+2 x_m\right)}{D}} v_1\bigg(\left(-2 x^2+2 x_m x+3 x_m^2\right) v_2^2-2 v_1 \left(L-x_m\right)\\
  &\quad\left(x+3 x_m\right) v_2+3 v_1^2 \left(L-x_m\right){}^2\bigg)\bigg)\bigg]\\
\end{split}
\label{Model3_Second_m_left}
\ee

\be
\begin{split}
\langle T\rangle_{+}
&=\frac{\left(e^{\frac{v_2 x_m}{D}}-e^{\frac{v_2 \left(L+x_m-x\right)}{D}}\right) \left(e^{\frac{v_1 x_m}{D}} \left((v_1+v_2)(D-v_1x_m)+L v_1^2\right)-D \left(v_1+v_2\right)\right)}{v_1 v_2\left(e^{\frac{L v_2}{D}} \left(\left(v_1+v_2\right) e^{\frac{v_1 x_m}{D}}-v_2\right)-v_1 e^{\frac{\left(v_1+v_2\right) x_m}{D}}\right)}\\
&\quad+\frac{L-x}{v_2}
\end{split}
\label{Model3_First_m_right}
\ee

\be
\begin{split}
  \langle T^2 \rangle_{+}
  &=\bigg[\frac{e^{-\frac{(L+x) v_2}{D}} }{v_1^3 v_2^3 \left(e^{\frac{L v_2}{D}} \left(v_2-\left(v_1+v_2\right) e^{\frac{v_1 x_m}{D}}\right)+v_1 e^{\frac{\left(v_1+v_2\right) x_m}{D}}\right){}^2}\bigg]\nonumber\\
  &\bigg[-2 \left(v_1+v_2\right) \bigg(e^{\frac{2 v_2 \left(L+x_m\right)}{D}} v_1 v_2 \left(v_1+v_2\right)-e^{\frac{v_2 \left(L+x+2 x_m\right)}{D}} v_1 v_2 \left(v_1+v_2\right)\nonumber\\
  & +e^{\frac{2 v_1 x_m+v_2 \left(3 L+x_m\right)}{D}} \left(v_1^2+v_2^2\right) \left(v_1+v_2\right)-e^{\frac{2 v_1 x_m+v_2 \left(2 L+x+x_m\right)}{D}} \left(v_1^2+v_2^2\right) \left(v_1+v_2\right)+v_2 \nonumber\\
  &e^{\frac{v_2 \left(3 L+x_m\right)}{D}} \left(2 v_1^2-2 v_2 v_1+v_2^2\right)-e^{\frac{v_2 \left(2 L+x+x_m\right)}{D}} v_2 \left(2 v_1^2-2 v_2 v_1+v_2^2\right)+e^{\frac{v_1 x_m+2 v_2 \left(L+x_m\right)}{D}} v_1\nonumber\\
  &\left(v_1^2-3 v_2 v_1+v_2^2\right)-e^{\frac{v_1 x_m+v_2 \left(L+x+2 x_m\right)}{D}} v_1 \left(v_1^2-3 v_2 v_1+v_2^2\right)+v_1 \left(v_1^2-2 v_2 v_1+2 v_2^2\right)\nonumber\\
  &e^{\frac{(L+x) v_2+2 \left(v_1+v_2\right) x_m}{D}} -e^{\frac{2 \left(v_1 x_m+v_2 \left(L+x_m\right)\right)}{D}} v_1 \left(v_1^2-2 v_2 v_1+2 v_2^2\right)-\left(v_1^3+3 v_2 v_1^2-v_2^2 v_1+2 v_2^3\right)\nonumber\\
  &e^{\frac{3 L v_2+\left(v_1+v_2\right) x_m}{D}} +e^{\frac{v_1 x_m+v_2 \left(2 L+x+x_m\right)}{D}} \left(v_1^3+3 v_2 v_1^2-v_2^2 v_1+2 v_2^3\right)\bigg)D^2+2 v_1 \bigg(e^{\frac{(3 L+x) v_2}{D}} (L-x)\nonumber\\
  &v_2^2 v_1^2+e^{\frac{(3 L+x) v_2+2 v_1 x_m}{D}} (L-x) \left(v_1+v_2\right){}^2 v_1^2-2 e^{\frac{(3 L+x) v_2+v_1 x_m}{D}} (L-x) v_2 \left(v_1+v_2\right) v_1^2\nonumber\\
  &+e^{\frac{(L+x) v_2+2 \left(v_1+v_2\right) x_m}{D}} \left(\left(x_m-x\right) v_1^2+v_2 \left(x-x_m\right) v_1+v_2^2 \left(x-2 x_m\right)\right) v_1^2+e^{\frac{2 \left(v_1 x_m+v_2 \left(L+x_m\right)\right)}{D}} \nonumber\\
  &\left(\left(L-x_m\right) v_1^2+v_2 \left(-2 L+x+x_m\right) v_1+v_2^2 \left(-2 L+x+2 x_m\right)\right) v_1^2+e^{\frac{v_2 \left(2 L+x+x_m\right)}{D}} v_2^2 \nonumber\\
  &\left(v_1+v_2\right) \left(3 L-x-2 x_m\right) v_1-e^{\frac{v_2 \left(3 L+x_m\right)}{D}} v_2^2 \left(v_1+v_2\right) \left(L+x-2 x_m\right) v_1+e^{\frac{v_1 x_m+v_2 \left(L+x+2 x_m\right)}{D}}\nonumber\\
  &v_2 \left(v_1+v_2\right) \left(3 v_2 x_m+v_1 \left(x_m-x\right)\right) v_1- v_2 \left(v_1+v_2\right) \left(3 v_2 x_m+v_1 \left(-2 L+x+x_m\right)\right) v_1\nonumber\\
  &e^{\frac{v_1 x_m+2 v_2 \left(L+x_m\right)}{D}}-e^{\frac{2 v_1 x_m+v_2 \left(3 L+x_m\right)}{D}} \left(v_1+v_2\right) \bigg(\left(L-x_m\right) v_1^3+v_2 \left(2 L+x-3 x_m\right) v_1^2-v_2^2\nonumber\\
  &\left(L-x+x_m\right) v_1+v_2^3 x_m\bigg)+e^{\frac{2 v_1 x_m+v_2 \left(2 L+x+x_m\right)}{D}} \left(v_1+v_2\right) \bigg(-\left(L-2 x+x_m\right) v_1^3+v_2 (4 L\nonumber\\
  & -x-3 x_m) v_1^2+v_2^2 \left(L-x-x_m\right) v_1+v_2^3 x_m\bigg)-e^{\frac{v_1 x_m+v_2 \left(2 L+x+x_m\right)}{D}} v_2 \bigg(\left(3 L+x-4 x_m\right) v_1^3\nonumber\\
  &+v_2 \left(8 L-3 x-4 x_m\right) v_1^2+v_2^2 \left(4 L-2 x+x_m\right) v_1+v_2^3 x_m\bigg)+e^{\frac{3 L v_2+\left(v_1+v_2\right) x_m}{D}} v_2 \bigg(L\nonumber\\
  &\left(3 v_1+2 v_2\right) v_1^2+\left(v_1+v_2\right) \left(\left(x-4 x_m\right) v_1^2+2 x v_2 v_1+v_2^2 x_m\right)\bigg)\bigg) D+v_1^2 v_2 \bigg(e^{\frac{(3 L+x) v_2}{D}}\nonumber\\
  &v_1 v_2^2 (L-x)^2+e^{\frac{(3 L+x) v_2+2 v_1 x_m}{D}} v_1 \left(v_1+v_2\right){}^2 (L-x)^2-2 e^{\frac{(3 L+x) v_2+v_1 x_m}{D}} v_1 v_2 \left(v_1+v_2\right)\nonumber\\
  &(L-x)^2+e^{\frac{(L+x) v_2+2 \left(v_1+v_2\right) x_m}{D}} v_1 \left(x v_1-\left(v_1+v_2\right) x_m\right){}^2+e^{\frac{2 v_1 x_m+v_2 \left(3 L+x_m\right)}{D}} \left(v_1+v_2\right)\nonumber\\
  &\left(v_2 x_m+v_1 \left(L+2 x-3 x_m\right)\right) \left(\left(v_1+v_2\right) x_m-L v_1\right)+e^{\frac{3 L v_2+\left(v_1+v_2\right) x_m}{D}} v_2 \left(\left(v_1+v_2\right) x_m-L v_1\right)\nonumber\\
  &\left(3 \left(v_1+v_2\right) x_m-(L+2 x) v_1\right)-e^{\frac{2 \left(v_1 x_m+v_2 \left(L+x_m\right)\right)}{D}} v_1 \left(L v_1-\left(v_1+v_2\right) x_m\right) (3 L v_1-2 x v_1\nonumber\\
  &-\left(v_1+v_2\right) x_m)-e^{\frac{v_1 x_m+v_2 \left(2 L+x+x_m\right)}{D}} v_2 \bigg(\left(3 L^2+2 x L-2 x^2\right) v_1^2-2 (4 L-x) \left(v_1+v_2\right)\nonumber\\
  & x_m v_1+3 \left(v_1+v_2\right){}^2 x_m^2\bigg)+e^{\frac{2 v_1 x_m+v_2 \left(2 L+x+x_m\right)}{D}} \left(v_1+v_2\right) \bigg(\bigg(3 L^2+2 x L-2 x^2+3 x_m^2\nonumber\\
    &+(2 x-8 L) x_m\bigg) v_1^2+2 v_2 x_m \left(-2 L+x+x_m\right) v_1-v_2^2 x_m^2\bigg)\bigg)\bigg]
  \end{split}
  \label{Model3_Second_m_right}
\ee


\texttt{MFPT with resetting}: We solve Eqn.(\ref{eq_lap_BFPE_1}, \ref{eq_lap_BFPE_2}) to find $\langle T_r \rangle_{-}=q_-(x,0)$ and $\langle T_r \rangle_{+}=q_+(x,0)$ as functions of $r$ by using matching conditions Eqn.(\ref{App_eq:Lap_survive_Match1}), (\ref{App_eq:Lap_survive_Match2}) at $x=x_m$ and boundary conditions $q_-(x,s)|_{x=0}=0$, $q_+(x,s)|_{x=L}=0$. The results are as follows:
\be
\begin{split}
\langle T_r \rangle_{-}
&= \bigg[\alpha ^- \left(e^{\alpha ^+ x_0}-1\right) \left(e^{\beta ^+ L+\left(\alpha ^-+\beta ^-\right) x_m}-e^{\beta ^- L+\left(\alpha ^-+\beta ^+\right) x_m}\right)+\beta ^-\bigg(e^{\beta ^+ L+\left(\alpha ^-+\beta ^-\right) x_m}\\
  &\quad-e^{\beta ^+ L+\left(\alpha ^-+\beta ^-\right) x_m+\alpha ^+ x_0}-e^{\beta ^+ L+\left(\alpha ^++\beta ^-\right) x_m}+e^{\beta ^+ L+\left(\alpha ^++\beta ^-\right) x_m+\alpha ^- x_0}\bigg) -\alpha ^+ (e^{\alpha ^- x_0}\\
  &\quad-1)e^{\beta ^+ L+\left(\alpha ^++\beta ^-\right) x_m}+\left(\alpha ^+-\beta ^+\right) \left(e^{\alpha ^- x_0}-1\right) e^{\beta ^- L+\left(\alpha ^++\beta ^+\right) x_m}+\beta ^+ e^{\left(\beta ^-+\beta ^+\right) x_m}\\
  &\quad(e^{\alpha ^- x_0}-e^{\alpha ^+ x_0})-\beta ^+ e^{\beta ^- L+\left(\alpha ^-+\beta ^+\right) x_m}(1-e^{\alpha ^+ x_0})+\beta ^- e^{\left(\beta ^-+\beta ^+\right) x_m}(e^{\alpha ^+ x_0}-e^{\alpha ^- x_0})\bigg]\bigg/\\
  & \quad\bigg[r\bigg( -\alpha ^- e^{\alpha ^+ x_0}\bigg(e^{\beta ^+ L+\left(\alpha ^-+\beta ^-\right) x_m}-e^{\beta ^- L+\left(\alpha ^-+\beta ^+\right) x_m}\bigg)+\beta ^- e^{\beta ^- x_m} \bigg(e^{\beta ^+ L} \big(e^{\alpha ^- x_m+\alpha ^+ x_0}\\
&\quad -e^{\alpha ^+ x_m+\alpha ^- x_0}\big)+\left(e^{\alpha ^- x_0}-e^{\alpha ^+ x_0}\right) e^{\beta ^+ x_m}\bigg)+\beta ^+ (e^{\alpha ^+ x_0}-e^{\alpha ^- x_0}-e^{\beta ^- L+\alpha ^+ x_0})\\
  &\quad e^{\left(\beta ^-+\beta ^+\right) x_m}+\alpha ^+ e^{\beta ^+ L+\left(\alpha ^++\beta ^-\right) x_m+\alpha ^- x_0}+\left(\beta ^+-\alpha ^+\right) e^{\beta ^- L+\left(\alpha ^++\beta ^+\right) x_m+\alpha ^- x_0}\bigg) \bigg]
\end{split}
\label{Model3_MFPT_w_reset_left}
\ee

\be
\begin{split}
\langle T_r \rangle_{+}
&= \bigg[\alpha ^- e^{\alpha ^- x_m} \bigg(-e^{\beta ^+ L+\alpha ^+ x_m+\beta ^- x_0}+e^{\beta ^- L+\alpha ^+ x_m+\beta ^+ x_0}+e^{\beta ^+ L+\beta ^- x_m}-e^{\beta ^- L+\beta ^+ x_m}\\
  &\quad-e^{\beta ^- x_m+\beta ^+ x_0}+e^{\beta ^+ x_m+\beta ^- x_0}\bigg)-\beta ^- \left(e^{\beta ^+ L}-e^{\beta ^+ x_0}\right) \left(e^{\left(\alpha ^-+\beta ^-\right) x_m}-e^{\left(\alpha ^++\beta ^-\right) x_m}\right)\\
  &\quad+\alpha ^+ \bigg(e^{\left(\alpha ^-+\alpha ^+\right) x_m}(e^{\beta ^+ L+\beta ^- x_0}-e^{\beta ^- L+\beta ^+ x_0})+e^{\left(\alpha ^++\beta ^-\right) x_m}(e^{\beta ^+ x_0}-e^{\beta ^+ L})\bigg)+(\alpha ^+\\
  &\quad-\beta ^+)\left(e^{\beta ^- L}-e^{\beta ^- x_0}\right) e^{\left(\alpha ^++\beta ^+\right) x_m}+\beta ^+ e^{\left(\alpha ^-+\beta ^+\right) x_m}(e^{\beta ^- L}-e^{\beta ^- x_0})\bigg]\bigg/\bigg[r\bigg(e^{\alpha ^+ x_m}\bigg(\\
  &\quad e^{\beta ^+ x_0} \bigg(e^{\beta ^- x_m}\left(\beta ^--\alpha ^+\right) +\left(\alpha ^+-\alpha ^-\right) e^{\beta ^- L+\alpha ^- x_m}\bigg)+e^{\beta ^- x_0} \bigg(\left(\alpha ^--\alpha ^+\right) e^{\beta ^+ L+\alpha ^- x_m}\\
  &\quad +\left(\alpha ^+-\beta ^+\right) e^{\beta ^+ x_m}\bigg)\bigg)   +e^{\alpha ^- x_m} \left(\left(\alpha ^--\beta ^-\right) e^{\beta ^- x_m+\beta ^+ x_0}+\left(\beta ^+-\alpha ^-\right) e^{\beta ^+ x_m+\beta ^- x_0}\right)\bigg)\bigg]\\
\end{split}
\label{Model3_MFPT_w_reset_right}
\ee

\par\texttt{Jump in ORR:} Here we show that the jump in $r_*$ at discontinuous transitions in Model-III may often be quite large such that analytical formula in Eq.(9) obtained for small order parameter approximation may not be very accurate.   In Fig.(\ref{fig:Model3_MFPT_vs_r}), we show $\langle T_r \rangle$ vs. $r$ for a certain DT --- while the exact value of $\Delta r_*=17.5$, the value from the formula in Eq.(9) is $10.032$ (marked as red dot on $r$-axis  in Fig.(\ref{fig:Model3_MFPT_vs_r})). So numerical method is better to find DT than approximate analytical formula in Eq.(9). 
\begin{figure}[ht!]
  \centering
  \includegraphics[width = 0.5\textwidth,height=0.23\textheight]{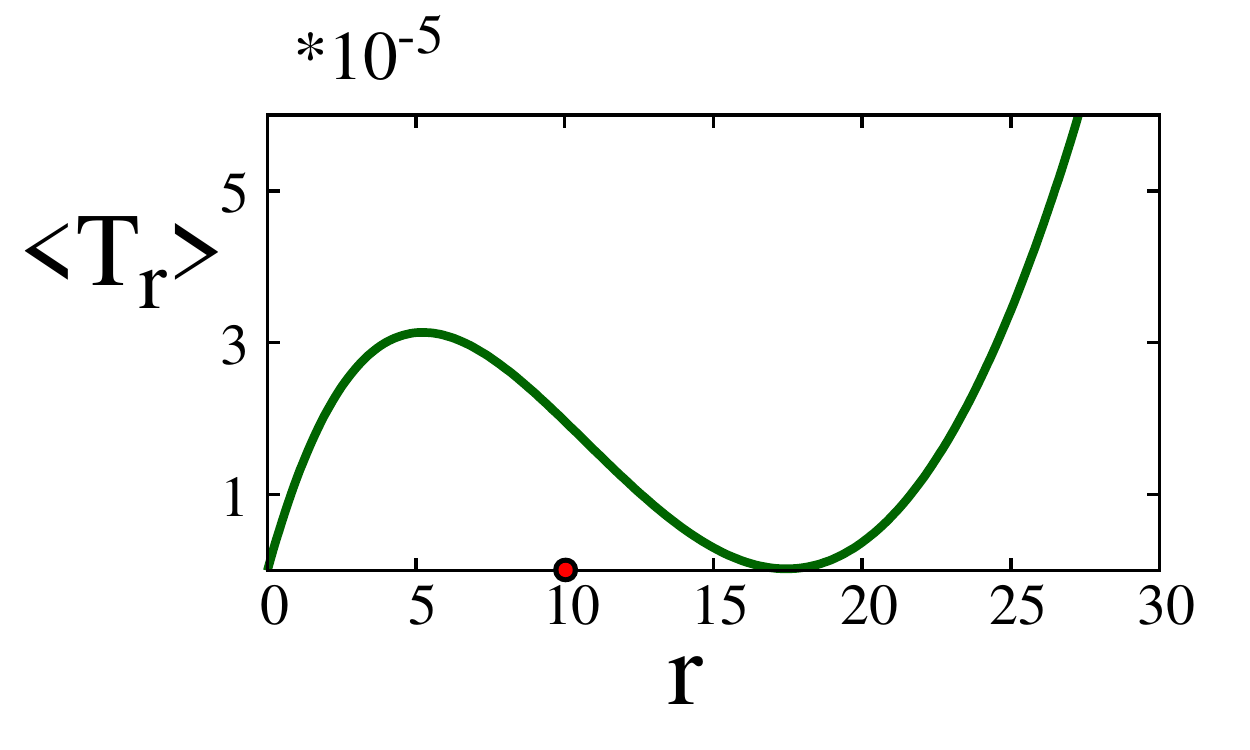}
    \caption{The figure shows MFPT with resetting plot at particular values of parameter when the discontinuous transition occur. The transition happen for $x_m=0.5$, $v_1=2$, $v_2=8.$ and $u_{0c}=0.19823$.}
    \label{fig:Model3_MFPT_vs_r}
\end{figure}

\par\texttt{Number of continuous transitions in Model-III for a fixed value of $v_1$ and $v_2\to\pm\infty$:} At $v_2\to\infty$ and any given $v_1$ for $x<x_m=0.5$, the condition of CT $\langle T^2 \rangle_{-}=2\langle T \rangle^2_{-}$ yields the following algebraic equation for $u_{0c}$: 
\begin{equation}
\begin{split}
 e^{v_1u_{0c}}(3v_1u_{0c}-0.75v_1-1)(1-e^{0.5v_1})
  &=e^{v_1} u_{0c}(v_1u_{0c}+2)-0.5v_1e^{2v_1u_{0c}}\\
 &\quad+e^{0.5v_1}(v_1(2u_{0c}^2-u_{0c}+0.75)-4u_{0c}+1)\\
  &\quad+ 2u_{0c}-1-u_{0c}(0.5-u_{0c})^2-2e^{v_1}v_1u_{0c}^2
\end{split}
\label{M3_v2_infty_left}
\end{equation}
The above equation has three real solutions of $u_{0c}$ indicating three CTs. For $x>x_m=0.5$ the corresponding equation obtained from $\langle T^2 \rangle_{+}=2\langle T \rangle^2_{+}$ is:
\be
u_{0c}-1=0
\label{M3_v2_infty_right}
\ee

which leads to the solution of $u_{0c} = 1$. Thus in total for the limit $v_2\to\infty$ there are four CTs.

For $v_2\to-\infty$ the transition condition is $\langle T^2 \rangle_{-}=2\langle T \rangle^2_{-}$ for $x<x_m=0.5$ and leads to:
\begin{equation} 
\begin{split}
e^{v_1u_{0c}}(e^{0.5v_1}(2-v_1(3u_{0c}-1))-1)
 &=e^{v_1}v_1u_{0c}(1-0.5v_1u_{0c})\\
 & \quad+e^{0.5v_1}(v_1(1-u_{0c})+2)-e^{2v_1u_{0c}}
\end{split}
\label{M3_v2_n_infty_left}
\end{equation}
The above equation has a single real solution of $u_{0c}$. For $x>x_m=0.5$ the condition for CT, $\langle T^2 \rangle_{+}=2\langle T \rangle^2_{+}$, is never satisfied as $\langle T^2 \rangle_{+}\gg 2\langle T \rangle^2_{+}$ even for finite $v_2<0$. We know that in this region there is a DT but no CT. Thus in the limit $v_2\to-\infty$ there is only one CT.
\end{widetext}
\bibliography{article_3_v1}
\end{document}